\documentclass[longauth]{aaEC}

\usepackage{graphicx}
\usepackage{natbib}
\usepackage{scalerel}

\usepackage[table]{xcolor}

\usepackage{txfonts}
\usepackage[pdfencoding=auto,psdextra]{hyperref}
\hypersetup{
    colorlinks=true,
    linkcolor=blue,
    filecolor=magenta,      
    urlcolor=blue,
    citecolor=blue
}
\makeatletter
\renewcommand*\aa@pageof{, page \thepage{} of \pageref*{LastPage}}
\makeatother

\usepackage{orcidlink} 
\newcommand{\orcid}[1]{\orcidlink{#1}}

\usepackage[utf8]{inputenc}

\usepackage[switch, modulo]{lineno}

\usepackage{euclid}

\begin{document}
%
%
\title{\Euclid preparation}
\subtitle{LVIII. Detecting extragalactic globular clusters in the Euclid survey}    

\author{Euclid Collaboration: K.~Voggel\orcid{0000-0001-6215-0950}\thanks{\email{karina.voggel@astro.unistra.fr}}\inst{\ref{aff1}}
\and A.~Lan\c{c}on\orcid{0000-0002-7214-8296}\inst{\ref{aff1}}
\and T.~Saifollahi\orcid{0000-0002-9554-7660}\inst{\ref{aff2},\ref{aff1}}
\and S.~S.~Larsen\orcid{0000-0003-0069-1203}\inst{\ref{aff3}}
\and M.~Cantiello\orcid{0000-0003-2072-384X}\inst{\ref{aff4}}
\and M.~Rejkuba\orcid{0000-0002-6577-2787}\inst{\ref{aff5}}
\and J.-C.~Cuillandre\orcid{0000-0002-3263-8645}\inst{\ref{aff6}}
\and P.~Hudelot\inst{\ref{aff7}}
\and A.~A.~Nucita\inst{\ref{aff8},\ref{aff9},\ref{aff10}}
\and M.~Urbano\orcid{0000-0001-5640-0650}\inst{\ref{aff1}}
\and E.~Romelli\orcid{0000-0003-3069-9222}\inst{\ref{aff11}}
\and M.~A.~Raj\orcid{0000-0002-8374-0340}\inst{\ref{aff2}}
\and M.~Schirmer\orcid{0000-0003-2568-9994}\inst{\ref{aff12}}
\and C.~Tortora\orcid{0000-0001-7958-6531}\inst{\ref{aff13}}
\and Abdurro'uf\inst{\ref{aff14}}
\and F.~Annibali\inst{\ref{aff15}}
\and M.~Baes\orcid{0000-0002-3930-2757}\inst{\ref{aff16}}
\and P.~Boldrini\inst{\ref{aff7}}
\and R.~Cabanac\orcid{0000-0001-6679-2600}\inst{\ref{aff17}}
\and D.~Carollo\orcid{0000-0002-0005-5787}\inst{\ref{aff11}}
\and C.~J.~Conselice\orcid{0000-0003-1949-7638}\inst{\ref{aff18}}
\and P.-A.~Duc\orcid{0000-0003-3343-6284}\inst{\ref{aff1}}
\and A.~M.~N.~Ferguson\inst{\ref{aff19}}
\and L.~K.~Hunt\orcid{0000-0001-9162-2371}\inst{\ref{aff20}}
\and J.~H.~Knapen\orcid{0000-0003-1643-0024}\inst{\ref{aff21},\ref{aff22}}
\and P.~Lonare\orcid{0009-0000-0028-0493}\inst{\ref{aff23},\ref{aff4}}
\and F.~R.~Marleau\inst{\ref{aff24}}
\and M.~Paolillo\orcid{0000-0003-4210-7693}\inst{\ref{aff25},\ref{aff13},\ref{aff26}}
\and M.~Poulain\orcid{0000-0002-7664-4510}\inst{\ref{aff27}}
\and R.~S\'anchez-Janssen\orcid{0000-0003-4945-0056}\inst{\ref{aff28}}
\and E.~Sola\orcid{0000-0002-2814-3578}\inst{\ref{aff29}}
\and S.~Andreon\orcid{0000-0002-2041-8784}\inst{\ref{aff30}}
\and N.~Auricchio\orcid{0000-0003-4444-8651}\inst{\ref{aff15}}
\and C.~Baccigalupi\orcid{0000-0002-8211-1630}\inst{\ref{aff31},\ref{aff11},\ref{aff32},\ref{aff33}}
\and M.~Baldi\orcid{0000-0003-4145-1943}\inst{\ref{aff34},\ref{aff15},\ref{aff35}}
\and S.~Bardelli\orcid{0000-0002-8900-0298}\inst{\ref{aff15}}
\and C.~Bodendorf\inst{\ref{aff36}}
\and D.~Bonino\orcid{0000-0002-3336-9977}\inst{\ref{aff37}}
\and E.~Branchini\orcid{0000-0002-0808-6908}\inst{\ref{aff38},\ref{aff39},\ref{aff30}}
\and M.~Brescia\orcid{0000-0001-9506-5680}\inst{\ref{aff40},\ref{aff13},\ref{aff26}}
\and J.~Brinchmann\orcid{0000-0003-4359-8797}\inst{\ref{aff41}}
\and S.~Camera\orcid{0000-0003-3399-3574}\inst{\ref{aff42},\ref{aff43},\ref{aff37}}
\and V.~Capobianco\orcid{0000-0002-3309-7692}\inst{\ref{aff37}}
\and C.~Carbone\orcid{0000-0003-0125-3563}\inst{\ref{aff44}}
\and R.~G.~Carlberg\inst{\ref{aff45}}
\and J.~Carretero\orcid{0000-0002-3130-0204}\inst{\ref{aff46},\ref{aff47}}
\and S.~Casas\orcid{0000-0002-4751-5138}\inst{\ref{aff48}}
\and M.~Castellano\orcid{0000-0001-9875-8263}\inst{\ref{aff49}}
\and G.~Castignani\orcid{0000-0001-6831-0687}\inst{\ref{aff15}}
\and S.~Cavuoti\orcid{0000-0002-3787-4196}\inst{\ref{aff13},\ref{aff26}}
\and A.~Cimatti\inst{\ref{aff50}}
\and C.~Colodro-Conde\inst{\ref{aff21}}
\and G.~Congedo\orcid{0000-0003-2508-0046}\inst{\ref{aff19}}
\and L.~Conversi\orcid{0000-0002-6710-8476}\inst{\ref{aff51},\ref{aff52}}
\and Y.~Copin\orcid{0000-0002-5317-7518}\inst{\ref{aff53}}
\and F.~Courbin\orcid{0000-0003-0758-6510}\inst{\ref{aff54}}
\and H.~M.~Courtois\orcid{0000-0003-0509-1776}\inst{\ref{aff55}}
\and M.~Cropper\orcid{0000-0003-4571-9468}\inst{\ref{aff56}}
\and A.~Da~Silva\orcid{0000-0002-6385-1609}\inst{\ref{aff57},\ref{aff58}}
\and H.~Degaudenzi\orcid{0000-0002-5887-6799}\inst{\ref{aff59}}
\and G.~De~Lucia\orcid{0000-0002-6220-9104}\inst{\ref{aff11}}
\and A.~M.~Di~Giorgio\orcid{0000-0002-4767-2360}\inst{\ref{aff60}}
\and J.~Dinis\inst{\ref{aff57},\ref{aff58}}
\and F.~Dubath\orcid{0000-0002-6533-2810}\inst{\ref{aff59}}
\and X.~Dupac\inst{\ref{aff52}}
\and S.~Dusini\orcid{0000-0002-1128-0664}\inst{\ref{aff61}}
\and M.~Farina\orcid{0000-0002-3089-7846}\inst{\ref{aff60}}
\and S.~Farrens\orcid{0000-0002-9594-9387}\inst{\ref{aff6}}
\and S.~Ferriol\inst{\ref{aff53}}
\and S.~Fotopoulou\orcid{0000-0002-9686-254X}\inst{\ref{aff62}}
\and M.~Frailis\orcid{0000-0002-7400-2135}\inst{\ref{aff11}}
\and E.~Franceschi\orcid{0000-0002-0585-6591}\inst{\ref{aff15}}
\and M.~Fumana\orcid{0000-0001-6787-5950}\inst{\ref{aff44}}
\and S.~Galeotta\orcid{0000-0002-3748-5115}\inst{\ref{aff11}}
\and K.~George\orcid{0000-0002-1734-8455}\inst{\ref{aff63}}
\and W.~Gillard\orcid{0000-0003-4744-9748}\inst{\ref{aff64}}
\and B.~Gillis\orcid{0000-0002-4478-1270}\inst{\ref{aff19}}
\and C.~Giocoli\orcid{0000-0002-9590-7961}\inst{\ref{aff15},\ref{aff65}}
\and P.~G\'omez-Alvarez\orcid{0000-0002-8594-5358}\inst{\ref{aff66},\ref{aff52}}
\and A.~Grazian\orcid{0000-0002-5688-0663}\inst{\ref{aff67}}
\and F.~Grupp\inst{\ref{aff36},\ref{aff63}}
\and S.~V.~H.~Haugan\orcid{0000-0001-9648-7260}\inst{\ref{aff68}}
\and H.~Hoekstra\orcid{0000-0002-0641-3231}\inst{\ref{aff69}}
\and W.~Holmes\inst{\ref{aff70}}
\and I.~Hook\orcid{0000-0002-2960-978X}\inst{\ref{aff71}}
\and F.~Hormuth\inst{\ref{aff72}}
\and A.~Hornstrup\orcid{0000-0002-3363-0936}\inst{\ref{aff73},\ref{aff74}}
\and K.~Jahnke\orcid{0000-0003-3804-2137}\inst{\ref{aff12}}
\and E.~Keih\"anen\orcid{0000-0003-1804-7715}\inst{\ref{aff75}}
\and S.~Kermiche\orcid{0000-0002-0302-5735}\inst{\ref{aff64}}
\and A.~Kiessling\orcid{0000-0002-2590-1273}\inst{\ref{aff70}}
\and M.~Kilbinger\orcid{0000-0001-9513-7138}\inst{\ref{aff6}}
\and R.~Kohley\inst{\ref{aff52}}
\and B.~Kubik\orcid{0009-0006-5823-4880}\inst{\ref{aff53}}
\and M.~K\"ummel\orcid{0000-0003-2791-2117}\inst{\ref{aff63}}
\and M.~Kunz\orcid{0000-0002-3052-7394}\inst{\ref{aff76}}
\and H.~Kurki-Suonio\orcid{0000-0002-4618-3063}\inst{\ref{aff77},\ref{aff78}}
\and R.~Laureijs\inst{\ref{aff79}}
\and P.~Liebing\inst{\ref{aff56}}
\and S.~Ligori\orcid{0000-0003-4172-4606}\inst{\ref{aff37}}
\and P.~B.~Lilje\orcid{0000-0003-4324-7794}\inst{\ref{aff68}}
\and V.~Lindholm\orcid{0000-0003-2317-5471}\inst{\ref{aff77},\ref{aff78}}
\and I.~Lloro\inst{\ref{aff80}}
\and D.~Maino\inst{\ref{aff81},\ref{aff44},\ref{aff82}}
\and E.~Maiorano\orcid{0000-0003-2593-4355}\inst{\ref{aff15}}
\and O.~Mansutti\orcid{0000-0001-5758-4658}\inst{\ref{aff11}}
\and O.~Marggraf\orcid{0000-0001-7242-3852}\inst{\ref{aff83}}
\and K.~Markovic\orcid{0000-0001-6764-073X}\inst{\ref{aff70}}
\and M.~Martinelli\orcid{0000-0002-6943-7732}\inst{\ref{aff49},\ref{aff84}}
\and N.~Martinet\orcid{0000-0003-2786-7790}\inst{\ref{aff85}}
\and F.~Marulli\orcid{0000-0002-8850-0303}\inst{\ref{aff86},\ref{aff15},\ref{aff35}}
\and R.~Massey\orcid{0000-0002-6085-3780}\inst{\ref{aff87},\ref{aff88}}
\and S.~Maurogordato\inst{\ref{aff89}}
\and E.~Medinaceli\orcid{0000-0002-4040-7783}\inst{\ref{aff15}}
\and S.~Mei\orcid{0000-0002-2849-559X}\inst{\ref{aff90}}
\and Y.~Mellier\inst{\ref{aff91},\ref{aff7}}
\and M.~Meneghetti\orcid{0000-0003-1225-7084}\inst{\ref{aff15},\ref{aff35}}
\and E.~Merlin\orcid{0000-0001-6870-8900}\inst{\ref{aff49}}
\and G.~Meylan\inst{\ref{aff54}}
\and M.~Moresco\orcid{0000-0002-7616-7136}\inst{\ref{aff86},\ref{aff15}}
\and L.~Moscardini\orcid{0000-0002-3473-6716}\inst{\ref{aff86},\ref{aff15},\ref{aff35}}
\and E.~Munari\orcid{0000-0002-1751-5946}\inst{\ref{aff11},\ref{aff31}}
\and R.~Nakajima\inst{\ref{aff83}}
\and C.~Neissner\orcid{0000-0001-8524-4968}\inst{\ref{aff92},\ref{aff47}}
\and R.~C.~Nichol\inst{\ref{aff93}}
\and S.-M.~Niemi\inst{\ref{aff79}}
\and J.~W.~Nightingale\orcid{0000-0002-8987-7401}\inst{\ref{aff94},\ref{aff88}}
\and C.~Padilla\orcid{0000-0001-7951-0166}\inst{\ref{aff92}}
\and S.~Paltani\orcid{0000-0002-8108-9179}\inst{\ref{aff59}}
\and F.~Pasian\orcid{0000-0002-4869-3227}\inst{\ref{aff11}}
\and K.~Pedersen\inst{\ref{aff95}}
\and V.~Pettorino\inst{\ref{aff79}}
\and S.~Pires\orcid{0000-0002-0249-2104}\inst{\ref{aff6}}
\and G.~Polenta\orcid{0000-0003-4067-9196}\inst{\ref{aff96}}
\and M.~Poncet\inst{\ref{aff97}}
\and L.~A.~Popa\inst{\ref{aff98}}
\and L.~Pozzetti\orcid{0000-0001-7085-0412}\inst{\ref{aff15}}
\and F.~Raison\orcid{0000-0002-7819-6918}\inst{\ref{aff36}}
\and R.~Rebolo\inst{\ref{aff21},\ref{aff22}}
\and A.~Renzi\orcid{0000-0001-9856-1970}\inst{\ref{aff99},\ref{aff61}}
\and J.~Rhodes\inst{\ref{aff70}}
\and G.~Riccio\inst{\ref{aff13}}
\and M.~Roncarelli\orcid{0000-0001-9587-7822}\inst{\ref{aff15}}
\and E.~Rossetti\orcid{0000-0003-0238-4047}\inst{\ref{aff34}}
\and R.~Saglia\orcid{0000-0003-0378-7032}\inst{\ref{aff63},\ref{aff36}}
\and Z.~Sakr\orcid{0000-0002-4823-3757}\inst{\ref{aff100},\ref{aff17},\ref{aff101}}
\and D.~Sapone\orcid{0000-0001-7089-4503}\inst{\ref{aff102}}
\and B.~Sartoris\orcid{0000-0003-1337-5269}\inst{\ref{aff63},\ref{aff11}}
\and R.~Scaramella\orcid{0000-0003-2229-193X}\inst{\ref{aff49},\ref{aff84}}
\and P.~Schneider\orcid{0000-0001-8561-2679}\inst{\ref{aff83}}
\and T.~Schrabback\orcid{0000-0002-6987-7834}\inst{\ref{aff24}}
\and A.~Secroun\orcid{0000-0003-0505-3710}\inst{\ref{aff64}}
\and E.~Sefusatti\orcid{0000-0003-0473-1567}\inst{\ref{aff11},\ref{aff31},\ref{aff32}}
\and G.~Seidel\orcid{0000-0003-2907-353X}\inst{\ref{aff12}}
\and S.~Serrano\orcid{0000-0002-0211-2861}\inst{\ref{aff103},\ref{aff104},\ref{aff105}}
\and C.~Sirignano\orcid{0000-0002-0995-7146}\inst{\ref{aff99},\ref{aff61}}
\and G.~Sirri\orcid{0000-0003-2626-2853}\inst{\ref{aff35}}
\and L.~Stanco\orcid{0000-0002-9706-5104}\inst{\ref{aff61}}
\and J.~Steinwagner\inst{\ref{aff36}}
\and C.~Surace\orcid{0000-0003-2592-0113}\inst{\ref{aff85}}
\and P.~Tallada-Cresp\'{i}\orcid{0000-0002-1336-8328}\inst{\ref{aff46},\ref{aff47}}
\and H.~I.~Teplitz\orcid{0000-0002-7064-5424}\inst{\ref{aff106}}
\and I.~Tereno\inst{\ref{aff57},\ref{aff107}}
\and R.~Toledo-Moreo\orcid{0000-0002-2997-4859}\inst{\ref{aff108}}
\and F.~Torradeflot\orcid{0000-0003-1160-1517}\inst{\ref{aff47},\ref{aff46}}
\and I.~Tutusaus\orcid{0000-0002-3199-0399}\inst{\ref{aff17}}
\and E.~A.~Valentijn\inst{\ref{aff2}}
\and L.~Valenziano\orcid{0000-0002-1170-0104}\inst{\ref{aff15},\ref{aff109}}
\and T.~Vassallo\orcid{0000-0001-6512-6358}\inst{\ref{aff63},\ref{aff11}}
\and A.~Veropalumbo\orcid{0000-0003-2387-1194}\inst{\ref{aff30},\ref{aff39}}
\and Y.~Wang\orcid{0000-0002-4749-2984}\inst{\ref{aff106}}
\and J.~Weller\orcid{0000-0002-8282-2010}\inst{\ref{aff63},\ref{aff36}}
\and G.~Zamorani\orcid{0000-0002-2318-301X}\inst{\ref{aff15}}
\and E.~Zucca\orcid{0000-0002-5845-8132}\inst{\ref{aff15}}
\and A.~Biviano\orcid{0000-0002-0857-0732}\inst{\ref{aff11},\ref{aff31}}
\and M.~Bolzonella\orcid{0000-0003-3278-4607}\inst{\ref{aff15}}
\and E.~Bozzo\orcid{0000-0002-8201-1525}\inst{\ref{aff59}}
\and C.~Burigana\orcid{0000-0002-3005-5796}\inst{\ref{aff110},\ref{aff109}}
\and M.~Calabrese\orcid{0000-0002-2637-2422}\inst{\ref{aff111},\ref{aff44}}
\and D.~Di~Ferdinando\inst{\ref{aff35}}
\and J.~A.~Escartin~Vigo\inst{\ref{aff36}}
\and R.~Farinelli\inst{\ref{aff15}}
\and J.~Gracia-Carpio\inst{\ref{aff36}}
\and N.~Mauri\orcid{0000-0001-8196-1548}\inst{\ref{aff50},\ref{aff35}}
\and V.~Scottez\inst{\ref{aff91},\ref{aff112}}
\and M.~Tenti\orcid{0000-0002-4254-5901}\inst{\ref{aff35}}
\and M.~Viel\orcid{0000-0002-2642-5707}\inst{\ref{aff31},\ref{aff11},\ref{aff33},\ref{aff32},\ref{aff113}}
\and M.~Wiesmann\orcid{0009-0000-8199-5860}\inst{\ref{aff68}}
\and Y.~Akrami\orcid{0000-0002-2407-7956}\inst{\ref{aff114},\ref{aff115}}
\and V.~Allevato\orcid{0000-0001-7232-5152}\inst{\ref{aff13}}
\and S.~Anselmi\orcid{0000-0002-3579-9583}\inst{\ref{aff61},\ref{aff99},\ref{aff116}}
\and M.~Ballardini\orcid{0000-0003-4481-3559}\inst{\ref{aff117},\ref{aff15},\ref{aff118}}
\and M.~Bethermin\orcid{0000-0002-3915-2015}\inst{\ref{aff1},\ref{aff85}}
\and A.~Blanchard\orcid{0000-0001-8555-9003}\inst{\ref{aff17}}
\and L.~Blot\orcid{0000-0002-9622-7167}\inst{\ref{aff119},\ref{aff116}}
\and S.~Borgani\orcid{0000-0001-6151-6439}\inst{\ref{aff120},\ref{aff31},\ref{aff11},\ref{aff32}}
\and A.~S.~Borlaff\orcid{0000-0003-3249-4431}\inst{\ref{aff121},\ref{aff122}}
\and S.~Bruton\orcid{0000-0002-6503-5218}\inst{\ref{aff123}}
\and A.~Calabro\orcid{0000-0003-2536-1614}\inst{\ref{aff49}}
\and G.~Canas-Herrera\orcid{0000-0003-2796-2149}\inst{\ref{aff79},\ref{aff124}}
\and A.~Cappi\inst{\ref{aff15},\ref{aff89}}
\and C.~S.~Carvalho\inst{\ref{aff107}}
\and T.~Castro\orcid{0000-0002-6292-3228}\inst{\ref{aff11},\ref{aff32},\ref{aff31},\ref{aff113}}
\and K.~C.~Chambers\orcid{0000-0001-6965-7789}\inst{\ref{aff125}}
\and S.~Contarini\orcid{0000-0002-9843-723X}\inst{\ref{aff36},\ref{aff86}}
\and A.~R.~Cooray\orcid{0000-0002-3892-0190}\inst{\ref{aff126}}
\and B.~De~Caro\inst{\ref{aff61},\ref{aff99}}
\and G.~Desprez\inst{\ref{aff127}}
\and A.~D\'iaz-S\'anchez\orcid{0000-0003-0748-4768}\inst{\ref{aff128}}
\and S.~Di~Domizio\orcid{0000-0003-2863-5895}\inst{\ref{aff38},\ref{aff39}}
\and H.~Dole\orcid{0000-0002-9767-3839}\inst{\ref{aff129}}
\and S.~Escoffier\orcid{0000-0002-2847-7498}\inst{\ref{aff64}}
\and I.~Ferrero\orcid{0000-0002-1295-1132}\inst{\ref{aff68}}
\and F.~Finelli\orcid{0000-0002-6694-3269}\inst{\ref{aff15},\ref{aff109}}
\and F.~Fornari\orcid{0000-0003-2979-6738}\inst{\ref{aff109}}
\and L.~Gabarra\orcid{0000-0002-8486-8856}\inst{\ref{aff130}}
\and K.~Ganga\orcid{0000-0001-8159-8208}\inst{\ref{aff90}}
\and J.~Garc\'ia-Bellido\orcid{0000-0002-9370-8360}\inst{\ref{aff114}}
\and V.~Gautard\inst{\ref{aff131}}
\and E.~Gaztanaga\orcid{0000-0001-9632-0815}\inst{\ref{aff104},\ref{aff103},\ref{aff132}}
\and F.~Giacomini\orcid{0000-0002-3129-2814}\inst{\ref{aff35}}
\and G.~Gozaliasl\orcid{0000-0002-0236-919X}\inst{\ref{aff133},\ref{aff77}}
\and A.~Hall\orcid{0000-0002-3139-8651}\inst{\ref{aff19}}
\and H.~Hildebrandt\orcid{0000-0002-9814-3338}\inst{\ref{aff134}}
\and J.~Hjorth\orcid{0000-0002-4571-2306}\inst{\ref{aff135}}
\and O.~Ilbert\orcid{0000-0002-7303-4397}\inst{\ref{aff85}}
\and J.~J.~E.~Kajava\orcid{0000-0002-3010-8333}\inst{\ref{aff136},\ref{aff137}}
\and V.~Kansal\orcid{0000-0002-4008-6078}\inst{\ref{aff138},\ref{aff139}}
\and D.~Karagiannis\orcid{0000-0002-4927-0816}\inst{\ref{aff140},\ref{aff141}}
\and C.~C.~Kirkpatrick\inst{\ref{aff75}}
\and L.~Legrand\orcid{0000-0003-0610-5252}\inst{\ref{aff142}}
\and G.~Libet\inst{\ref{aff97}}
\and A.~Loureiro\orcid{0000-0002-4371-0876}\inst{\ref{aff143},\ref{aff144}}
\and J.~Macias-Perez\orcid{0000-0002-5385-2763}\inst{\ref{aff145}}
\and G.~Maggio\orcid{0000-0003-4020-4836}\inst{\ref{aff11}}
\and M.~Magliocchetti\orcid{0000-0001-9158-4838}\inst{\ref{aff60}}
\and F.~Mannucci\orcid{0000-0002-4803-2381}\inst{\ref{aff20}}
\and R.~Maoli\orcid{0000-0002-6065-3025}\inst{\ref{aff146},\ref{aff49}}
\and C.~J.~A.~P.~Martins\orcid{0000-0002-4886-9261}\inst{\ref{aff147},\ref{aff41}}
\and S.~Matthew\inst{\ref{aff19}}
\and L.~Maurin\orcid{0000-0002-8406-0857}\inst{\ref{aff129}}
\and R.~B.~Metcalf\orcid{0000-0003-3167-2574}\inst{\ref{aff86},\ref{aff15}}
\and P.~Monaco\orcid{0000-0003-2083-7564}\inst{\ref{aff120},\ref{aff11},\ref{aff32},\ref{aff31}}
\and C.~Moretti\orcid{0000-0003-3314-8936}\inst{\ref{aff33},\ref{aff113},\ref{aff11},\ref{aff31},\ref{aff32}}
\and G.~Morgante\inst{\ref{aff15}}
\and Nicholas~A.~Walton\orcid{0000-0003-3983-8778}\inst{\ref{aff29}}
\and L.~Patrizii\inst{\ref{aff35}}
\and A.~Pezzotta\orcid{0000-0003-0726-2268}\inst{\ref{aff36}}
\and M.~P\"ontinen\orcid{0000-0001-5442-2530}\inst{\ref{aff77}}
\and V.~Popa\inst{\ref{aff98}}
\and C.~Porciani\orcid{0000-0002-7797-2508}\inst{\ref{aff83}}
\and D.~Potter\orcid{0000-0002-0757-5195}\inst{\ref{aff148}}
\and P.~Reimberg\orcid{0000-0003-3410-0280}\inst{\ref{aff91}}
\and I.~Risso\orcid{0000-0003-2525-7761}\inst{\ref{aff149}}
\and P.-F.~Rocci\inst{\ref{aff129}}
\and M.~Sahl\'en\orcid{0000-0003-0973-4804}\inst{\ref{aff150}}
\and A.~Schneider\orcid{0000-0001-7055-8104}\inst{\ref{aff148}}
\and M.~Sereno\orcid{0000-0003-0302-0325}\inst{\ref{aff15},\ref{aff35}}
\and P.~Simon\inst{\ref{aff83}}
\and A.~Spurio~Mancini\orcid{0000-0001-5698-0990}\inst{\ref{aff151},\ref{aff56}}
\and G.~Testera\inst{\ref{aff39}}
\and R.~Teyssier\orcid{0000-0001-7689-0933}\inst{\ref{aff152}}
\and S.~Toft\orcid{0000-0003-3631-7176}\inst{\ref{aff74},\ref{aff153},\ref{aff154}}
\and S.~Tosi\orcid{0000-0002-7275-9193}\inst{\ref{aff38},\ref{aff39},\ref{aff30}}
\and A.~Troja\orcid{0000-0003-0239-4595}\inst{\ref{aff99},\ref{aff61}}
\and M.~Tucci\inst{\ref{aff59}}
\and J.~Valiviita\orcid{0000-0001-6225-3693}\inst{\ref{aff77},\ref{aff78}}
\and D.~Vergani\orcid{0000-0003-0898-2216}\inst{\ref{aff15}}
\and G.~Verza\orcid{0000-0002-1886-8348}\inst{\ref{aff155},\ref{aff156}}
\and I.~A.~Zinchenko\inst{\ref{aff63}}
\and G.~A.~Mamon\orcid{0000-0001-8956-5953}\inst{\ref{aff7},\ref{aff91}}
\and D.~Scott\orcid{0000-0002-6878-9840}\inst{\ref{aff157}}}
										   
\institute{Universit\'e de Strasbourg, CNRS, Observatoire astronomique de Strasbourg, UMR 7550, 67000 Strasbourg, France\label{aff1}
\and
Kapteyn Astronomical Institute, University of Groningen, PO Box 800, 9700 AV Groningen, The Netherlands\label{aff2}
\and
Department of Astrophysics/IMAPP, Radboud University, PO Box 9010, 6500 GL Nijmegen, The Netherlands\label{aff3}
\and
INAF - Osservatorio Astronomico d'Abruzzo, Via Maggini, 64100, Teramo, Italy\label{aff4}
\and
European Southern Observatory, Karl-Schwarzschild-Str.~2, 85748 Garching, Germany\label{aff5}
\and
Universit\'e Paris-Saclay, Universit\'e Paris Cit\'e, CEA, CNRS, AIM, 91191, Gif-sur-Yvette, France\label{aff6}
\and
Institut d'Astrophysique de Paris, UMR 7095, CNRS, and Sorbonne Universit\'e, 98 bis boulevard Arago, 75014 Paris, France\label{aff7}
\and
Department of Mathematics and Physics E. De Giorgi, University of Salento, Via per Arnesano, CP-I93, 73100, Lecce, Italy\label{aff8}
\and
INAF-Sezione di Lecce, c/o Dipartimento Matematica e Fisica, Via per Arnesano, 73100, Lecce, Italy\label{aff9}
\and
INFN, Sezione di Lecce, Via per Arnesano, CP-193, 73100, Lecce, Italy\label{aff10}
\and
INAF-Osservatorio Astronomico di Trieste, Via G. B. Tiepolo 11, 34143 Trieste, Italy\label{aff11}
\and
Max-Planck-Institut f\"ur Astronomie, K\"onigstuhl 17, 69117 Heidelberg, Germany\label{aff12}
\and
INAF-Osservatorio Astronomico di Capodimonte, Via Moiariello 16, 80131 Napoli, Italy\label{aff13}
\and
Johns Hopkins University 3400 North Charles Street Baltimore, MD 21218, USA\label{aff14}
\and
INAF-Osservatorio di Astrofisica e Scienza dello Spazio di Bologna, Via Piero Gobetti 93/3, 40129 Bologna, Italy\label{aff15}
\and
Sterrenkundig Observatorium, Universiteit Gent, Krijgslaan 281 S9, 9000 Gent, Belgium\label{aff16}
\and
Institut de Recherche en Astrophysique et Plan\'etologie (IRAP), Universit\'e de Toulouse, CNRS, UPS, CNES, 14 Av. Edouard Belin, 31400 Toulouse, France\label{aff17}
\and
Jodrell Bank Centre for Astrophysics, Department of Physics and Astronomy, University of Manchester, Oxford Road, Manchester M13 9PL, UK\label{aff18}
\and
Institute for Astronomy, University of Edinburgh, Royal Observatory, Blackford Hill, Edinburgh EH9 3HJ, UK\label{aff19}
\and
INAF-Osservatorio Astrofisico di Arcetri, Largo E. Fermi 5, 50125, Firenze, Italy\label{aff20}
\and
Instituto de Astrof\'isica de Canarias, Calle V\'ia L\'actea s/n, 38204, San Crist\'obal de La Laguna, Tenerife, Spain\label{aff21}
\and
Departamento de Astrof\'isica, Universidad de La Laguna, 38206, La Laguna, Tenerife, Spain\label{aff22}
\and
Dipartimento di Fisica, Universit\`a di Roma Tor Vergata, Via della Ricerca Scientifica 1, Roma, Italy\label{aff23}
\and
Universit\"at Innsbruck, Institut f\"ur Astro- und Teilchenphysik, Technikerstr. 25/8, 6020 Innsbruck, Austria\label{aff24}
\and
Dipartimento di Fisica "E. Pancini", Universita degli Studi di Napoli Federico II, Via Cinthia 6, 80126, Napoli, Italy\label{aff25}
\and
INFN section of Naples, Via Cinthia 6, 80126, Napoli, Italy\label{aff26}
\and
Space physics and astronomy research unit, University of Oulu, Pentti Kaiteran katu 1, FI-90014 Oulu, Finland\label{aff27}
\and
UK Astronomy Technology Centre, Royal Observatory, Blackford Hill, Edinburgh EH9 3HJ, UK\label{aff28}
\and
Institute of Astronomy, University of Cambridge, Madingley Road, Cambridge CB3 0HA, UK\label{aff29}
\and
INAF-Osservatorio Astronomico di Brera, Via Brera 28, 20122 Milano, Italy\label{aff30}
\and
IFPU, Institute for Fundamental Physics of the Universe, via Beirut 2, 34151 Trieste, Italy\label{aff31}
\and
INFN, Sezione di Trieste, Via Valerio 2, 34127 Trieste TS, Italy\label{aff32}
\and
SISSA, International School for Advanced Studies, Via Bonomea 265, 34136 Trieste TS, Italy\label{aff33}
\and
Dipartimento di Fisica e Astronomia, Universit\`a di Bologna, Via Gobetti 93/2, 40129 Bologna, Italy\label{aff34}
\and
INFN-Sezione di Bologna, Viale Berti Pichat 6/2, 40127 Bologna, Italy\label{aff35}
\and
Max Planck Institute for Extraterrestrial Physics, Giessenbachstr. 1, 85748 Garching, Germany\label{aff36}
\and
INAF-Osservatorio Astrofisico di Torino, Via Osservatorio 20, 10025 Pino Torinese (TO), Italy\label{aff37}
\and
Dipartimento di Fisica, Universit\`a di Genova, Via Dodecaneso 33, 16146, Genova, Italy\label{aff38}
\and
INFN-Sezione di Genova, Via Dodecaneso 33, 16146, Genova, Italy\label{aff39}
\and
Department of Physics "E. Pancini", University Federico II, Via Cinthia 6, 80126, Napoli, Italy\label{aff40}
\and
Instituto de Astrof\'isica e Ci\^encias do Espa\c{c}o, Universidade do Porto, CAUP, Rua das Estrelas, PT4150-762 Porto, Portugal\label{aff41}
\and
Dipartimento di Fisica, Universit\`a degli Studi di Torino, Via P. Giuria 1, 10125 Torino, Italy\label{aff42}
\and
INFN-Sezione di Torino, Via P. Giuria 1, 10125 Torino, Italy\label{aff43}
\and
INAF-IASF Milano, Via Alfonso Corti 12, 20133 Milano, Italy\label{aff44}
\and
David A. Dunlap Department of Astronomy \& Astrophysics, University of Toronto, 50 St George Street, Toronto, Ontario M5S 3H4, Canada\label{aff45}
\and
Centro de Investigaciones Energ\'eticas, Medioambientales y Tecnol\'ogicas (CIEMAT), Avenida Complutense 40, 28040 Madrid, Spain\label{aff46}
\and
Port d'Informaci\'{o} Cient\'{i}fica, Campus UAB, C. Albareda s/n, 08193 Bellaterra (Barcelona), Spain\label{aff47}
\and
Institute for Theoretical Particle Physics and Cosmology (TTK), RWTH Aachen University, 52056 Aachen, Germany\label{aff48}
\and
INAF-Osservatorio Astronomico di Roma, Via Frascati 33, 00078 Monteporzio Catone, Italy\label{aff49}
\and
Dipartimento di Fisica e Astronomia "Augusto Righi" - Alma Mater Studiorum Universit\`a di Bologna, Viale Berti Pichat 6/2, 40127 Bologna, Italy\label{aff50}
\and
European Space Agency/ESRIN, Largo Galileo Galilei 1, 00044 Frascati, Roma, Italy\label{aff51}
\and
ESAC/ESA, Camino Bajo del Castillo, s/n., Urb. Villafranca del Castillo, 28692 Villanueva de la Ca\~nada, Madrid, Spain\label{aff52}
\and
Universit\'e Claude Bernard Lyon 1, CNRS/IN2P3, IP2I Lyon, UMR 5822, Villeurbanne, F-69100, France\label{aff53}
\and
Institute of Physics, Laboratory of Astrophysics, Ecole Polytechnique F\'ed\'erale de Lausanne (EPFL), Observatoire de Sauverny, 1290 Versoix, Switzerland\label{aff54}
\and
UCB Lyon 1, CNRS/IN2P3, IUF, IP2I Lyon, 4 rue Enrico Fermi, 69622 Villeurbanne, France\label{aff55}
\and
Mullard Space Science Laboratory, University College London, Holmbury St Mary, Dorking, Surrey RH5 6NT, UK\label{aff56}
\and
Departamento de F\'isica, Faculdade de Ci\^encias, Universidade de Lisboa, Edif\'icio C8, Campo Grande, PT1749-016 Lisboa, Portugal\label{aff57}
\and
Instituto de Astrof\'isica e Ci\^encias do Espa\c{c}o, Faculdade de Ci\^encias, Universidade de Lisboa, Campo Grande, 1749-016 Lisboa, Portugal\label{aff58}
\and
Department of Astronomy, University of Geneva, ch. d'Ecogia 16, 1290 Versoix, Switzerland\label{aff59}
\and
INAF-Istituto di Astrofisica e Planetologia Spaziali, via del Fosso del Cavaliere, 100, 00100 Roma, Italy\label{aff60}
\and
INFN-Padova, Via Marzolo 8, 35131 Padova, Italy\label{aff61}
\and
School of Physics, HH Wills Physics Laboratory, University of Bristol, Tyndall Avenue, Bristol, BS8 1TL, UK\label{aff62}
\and
Universit\"ats-Sternwarte M\"unchen, Fakult\"at f\"ur Physik, Ludwig-Maximilians-Universit\"at M\"unchen, Scheinerstrasse 1, 81679 M\"unchen, Germany\label{aff63}
\and
Aix-Marseille Universit\'e, CNRS/IN2P3, CPPM, Marseille, France\label{aff64}
\and
Istituto Nazionale di Fisica Nucleare, Sezione di Bologna, Via Irnerio 46, 40126 Bologna, Italy\label{aff65}
\and
FRACTAL S.L.N.E., calle Tulip\'an 2, Portal 13 1A, 28231, Las Rozas de Madrid, Spain\label{aff66}
\and
INAF-Osservatorio Astronomico di Padova, Via dell'Osservatorio 5, 35122 Padova, Italy\label{aff67}
\and
Institute of Theoretical Astrophysics, University of Oslo, P.O. Box 1029 Blindern, 0315 Oslo, Norway\label{aff68}
\and
Leiden Observatory, Leiden University, Einsteinweg 55, 2333 CC Leiden, The Netherlands\label{aff69}
\and
Jet Propulsion Laboratory, California Institute of Technology, 4800 Oak Grove Drive, Pasadena, CA, 91109, USA\label{aff70}
\and
Department of Physics, Lancaster University, Lancaster, LA1 4YB, UK\label{aff71}
\and
Felix Hormuth Engineering, Goethestr. 17, 69181 Leimen, Germany\label{aff72}
\and
Technical University of Denmark, Elektrovej 327, 2800 Kgs. Lyngby, Denmark\label{aff73}
\and
Cosmic Dawn Center (DAWN), Denmark\label{aff74}
\and
Department of Physics and Helsinki Institute of Physics, Gustaf H\"allstr\"omin katu 2, 00014 University of Helsinki, Finland\label{aff75}
\and
Universit\'e de Gen\`eve, D\'epartement de Physique Th\'eorique and Centre for Astroparticle Physics, 24 quai Ernest-Ansermet, CH-1211 Gen\`eve 4, Switzerland\label{aff76}
\and
Department of Physics, P.O. Box 64, 00014 University of Helsinki, Finland\label{aff77}
\and
Helsinki Institute of Physics, Gustaf H{\"a}llstr{\"o}min katu 2, University of Helsinki, Helsinki, Finland\label{aff78}
\and
European Space Agency/ESTEC, Keplerlaan 1, 2201 AZ Noordwijk, The Netherlands\label{aff79}
\and
NOVA optical infrared instrumentation group at ASTRON, Oude Hoogeveensedijk 4, 7991PD, Dwingeloo, The Netherlands\label{aff80}
\and
Dipartimento di Fisica "Aldo Pontremoli", Universit\`a degli Studi di Milano, Via Celoria 16, 20133 Milano, Italy\label{aff81}
\and
INFN-Sezione di Milano, Via Celoria 16, 20133 Milano, Italy\label{aff82}
\and
Universit\"at Bonn, Argelander-Institut f\"ur Astronomie, Auf dem H\"ugel 71, 53121 Bonn, Germany\label{aff83}
\and
INFN-Sezione di Roma, Piazzale Aldo Moro, 2 - c/o Dipartimento di Fisica, Edificio G. Marconi, 00185 Roma, Italy\label{aff84}
\and
Aix-Marseille Universit\'e, CNRS, CNES, LAM, Marseille, France\label{aff85}
\and
Dipartimento di Fisica e Astronomia "Augusto Righi" - Alma Mater Studiorum Universit\`a di Bologna, via Piero Gobetti 93/2, 40129 Bologna, Italy\label{aff86}
\and
Department of Physics, Centre for Extragalactic Astronomy, Durham University, South Road, DH1 3LE, UK\label{aff87}
\and
Department of Physics, Institute for Computational Cosmology, Durham University, South Road, DH1 3LE, UK\label{aff88}
\and
Universit\'e C\^{o}te d'Azur, Observatoire de la C\^{o}te d'Azur, CNRS, Laboratoire Lagrange, Bd de l'Observatoire, CS 34229, 06304 Nice cedex 4, France\label{aff89}
\and
Universit\'e Paris Cit\'e, CNRS, Astroparticule et Cosmologie, 75013 Paris, France\label{aff90}
\and
Institut d'Astrophysique de Paris, 98bis Boulevard Arago, 75014, Paris, France\label{aff91}
\and
Institut de F\'{i}sica d'Altes Energies (IFAE), The Barcelona Institute of Science and Technology, Campus UAB, 08193 Bellaterra (Barcelona), Spain\label{aff92}
\and
School of Mathematics and Physics, University of Surrey, Guildford, Surrey, GU2 7XH, UK\label{aff93}
\and
School of Mathematics, Statistics and Physics, Newcastle University, Herschel Building, Newcastle-upon-Tyne, NE1 7RU, UK\label{aff94}
\and
Department of Physics and Astronomy, University of Aarhus, Ny Munkegade 120, DK-8000 Aarhus C, Denmark\label{aff95}
\and
Space Science Data Center, Italian Space Agency, via del Politecnico snc, 00133 Roma, Italy\label{aff96}
\and
Centre National d'Etudes Spatiales -- Centre spatial de Toulouse, 18 avenue Edouard Belin, 31401 Toulouse Cedex 9, France\label{aff97}
\and
Institute of Space Science, Str. Atomistilor, nr. 409 M\u{a}gurele, Ilfov, 077125, Romania\label{aff98}
\and
Dipartimento di Fisica e Astronomia "G. Galilei", Universit\`a di Padova, Via Marzolo 8, 35131 Padova, Italy\label{aff99}
\and
Institut f\"ur Theoretische Physik, University of Heidelberg, Philosophenweg 16, 69120 Heidelberg, Germany\label{aff100}
\and
Universit\'e St Joseph; Faculty of Sciences, Beirut, Lebanon\label{aff101}
\and
Departamento de F\'isica, FCFM, Universidad de Chile, Blanco Encalada 2008, Santiago, Chile\label{aff102}
\and
Institut d'Estudis Espacials de Catalunya (IEEC),  Edifici RDIT, Campus UPC, 08860 Castelldefels, Barcelona, Spain\label{aff103}
\and
Institute of Space Sciences (ICE, CSIC), Campus UAB, Carrer de Can Magrans, s/n, 08193 Barcelona, Spain\label{aff104}
\and
Satlantis, University Science Park, Sede Bld 48940, Leioa-Bilbao, Spain\label{aff105}
\and
Infrared Processing and Analysis Center, California Institute of Technology, Pasadena, CA 91125, USA\label{aff106}
\and
Instituto de Astrof\'isica e Ci\^encias do Espa\c{c}o, Faculdade de Ci\^encias, Universidade de Lisboa, Tapada da Ajuda, 1349-018 Lisboa, Portugal\label{aff107}
\and
Universidad Polit\'ecnica de Cartagena, Departamento de Electr\'onica y Tecnolog\'ia de Computadoras,  Plaza del Hospital 1, 30202 Cartagena, Spain\label{aff108}
\and
INFN-Bologna, Via Irnerio 46, 40126 Bologna, Italy\label{aff109}
\and
INAF, Istituto di Radioastronomia, Via Piero Gobetti 101, 40129 Bologna, Italy\label{aff110}
\and
Astronomical Observatory of the Autonomous Region of the Aosta Valley (OAVdA), Loc. Lignan 39, I-11020, Nus (Aosta Valley), Italy\label{aff111}
\and
ICL, Junia, Universit\'e Catholique de Lille, LITL, 59000 Lille, France\label{aff112}
\and
ICSC - Centro Nazionale di Ricerca in High Performance Computing, Big Data e Quantum Computing, Via Magnanelli 2, Bologna, Italy\label{aff113}
\and
Instituto de F\'isica Te\'orica UAM-CSIC, Campus de Cantoblanco, 28049 Madrid, Spain\label{aff114}
\and
CERCA/ISO, Department of Physics, Case Western Reserve University, 10900 Euclid Avenue, Cleveland, OH 44106, USA\label{aff115}
\and
Laboratoire Univers et Th\'eorie, Observatoire de Paris, Universit\'e PSL, Universit\'e Paris Cit\'e, CNRS, 92190 Meudon, France\label{aff116}
\and
Dipartimento di Fisica e Scienze della Terra, Universit\`a degli Studi di Ferrara, Via Giuseppe Saragat 1, 44122 Ferrara, Italy\label{aff117}
\and
Istituto Nazionale di Fisica Nucleare, Sezione di Ferrara, Via Giuseppe Saragat 1, 44122 Ferrara, Italy\label{aff118}
\and
Kavli Institute for the Physics and Mathematics of the Universe (WPI), University of Tokyo, Kashiwa, Chiba 277-8583, Japan\label{aff119}
\and
Dipartimento di Fisica - Sezione di Astronomia, Universit\`a di Trieste, Via Tiepolo 11, 34131 Trieste, Italy\label{aff120}
\and
NASA Ames Research Center, Moffett Field, CA 94035, USA\label{aff121}
\and
Bay Area Environmental Research Institute, Moffett Field, California 94035, USA\label{aff122}
\and
Minnesota Institute for Astrophysics, University of Minnesota, 116 Church St SE, Minneapolis, MN 55455, USA\label{aff123}
\and
Institute Lorentz, Leiden University, Niels Bohrweg 2, 2333 CA Leiden, The Netherlands\label{aff124}
\and
Institute for Astronomy, University of Hawaii, 2680 Woodlawn Drive, Honolulu, HI 96822, USA\label{aff125}
\and
Department of Physics \& Astronomy, University of California Irvine, Irvine CA 92697, USA\label{aff126}
\and
Department of Astronomy \& Physics and Institute for Computational Astrophysics, Saint Mary's University, 923 Robie Street, Halifax, Nova Scotia, B3H 3C3, Canada\label{aff127}
\and
Departamento F\'isica Aplicada, Universidad Polit\'ecnica de Cartagena, Campus Muralla del Mar, 30202 Cartagena, Murcia, Spain\label{aff128}
\and
Universit\'e Paris-Saclay, CNRS, Institut d'astrophysique spatiale, 91405, Orsay, France\label{aff129}
\and
Department of Physics, Oxford University, Keble Road, Oxford OX1 3RH, UK\label{aff130}
\and
CEA Saclay, DFR/IRFU, Service d'Astrophysique, Bat. 709, 91191 Gif-sur-Yvette, France\label{aff131}
\and
Institute of Cosmology and Gravitation, University of Portsmouth, Portsmouth PO1 3FX, UK\label{aff132}
\and
Department of Computer Science, Aalto University, PO Box 15400, Espoo, FI-00 076, Finland\label{aff133}
\and
Ruhr University Bochum, Faculty of Physics and Astronomy, Astronomical Institute (AIRUB), German Centre for Cosmological Lensing (GCCL), 44780 Bochum, Germany\label{aff134}
\and
DARK, Niels Bohr Institute, University of Copenhagen, Jagtvej 155, 2200 Copenhagen, Denmark\label{aff135}
\and
Department of Physics and Astronomy, Vesilinnantie 5, 20014 University of Turku, Finland\label{aff136}
\and
Serco for European Space Agency (ESA), Camino bajo del Castillo, s/n, Urbanizacion Villafranca del Castillo, Villanueva de la Ca\~nada, 28692 Madrid, Spain\label{aff137}
\and
ARC Centre of Excellence for Dark Matter Particle Physics, Melbourne, Australia\label{aff138}
\and
Centre for Astrophysics \& Supercomputing, Swinburne University of Technology,  Hawthorn, Victoria 3122, Australia\label{aff139}
\and
School of Physics and Astronomy, Queen Mary University of London, Mile End Road, London E1 4NS, UK\label{aff140}
\and
Department of Physics and Astronomy, University of the Western Cape, Bellville, Cape Town, 7535, South Africa\label{aff141}
\and
ICTP South American Institute for Fundamental Research, Instituto de F\'{\i}sica Te\'orica, Universidade Estadual Paulista, S\~ao Paulo, Brazil\label{aff142}
\and
Oskar Klein Centre for Cosmoparticle Physics, Department of Physics, Stockholm University, Stockholm, SE-106 91, Sweden\label{aff143}
\and
Astrophysics Group, Blackett Laboratory, Imperial College London, London SW7 2AZ, UK\label{aff144}
\and
Univ. Grenoble Alpes, CNRS, Grenoble INP, LPSC-IN2P3, 53, Avenue des Martyrs, 38000, Grenoble, France\label{aff145}
\and
Dipartimento di Fisica, Sapienza Universit\`a di Roma, Piazzale Aldo Moro 2, 00185 Roma, Italy\label{aff146}
\and
Centro de Astrof\'{\i}sica da Universidade do Porto, Rua das Estrelas, 4150-762 Porto, Portugal\label{aff147}
\and
Department of Astrophysics, University of Zurich, Winterthurerstrasse 190, 8057 Zurich, Switzerland\label{aff148}
\and
Dipartimento di Fisica, Universit\`a degli studi di Genova, and INFN-Sezione di Genova, via Dodecaneso 33, 16146, Genova, Italy\label{aff149}
\and
Theoretical astrophysics, Department of Physics and Astronomy, Uppsala University, Box 515, 751 20 Uppsala, Sweden\label{aff150}
\and
Department of Physics, Royal Holloway, University of London, TW20 0EX, UK\label{aff151}
\and
Department of Astrophysical Sciences, Peyton Hall, Princeton University, Princeton, NJ 08544, USA\label{aff152}
\and
Cosmic Dawn Center (DAWN)\label{aff153}
\and
Niels Bohr Institute, University of Copenhagen, Jagtvej 128, 2200 Copenhagen, Denmark\label{aff154}
\and
Center for Cosmology and Particle Physics, Department of Physics, New York University, New York, NY 10003, USA\label{aff155}
\and
Center for Computational Astrophysics, Flatiron Institute, 162 5th Avenue, 10010, New York, NY, USA\label{aff156}
\and
Department of Physics and Astronomy, University of British Columbia, Vancouver, BC V6T 1Z1, Canada\label{aff157}}

 \date{\bf \today}

%
%
 \abstract{
Extragalactic globular clusters (EGCs) are an abundant and powerful tracer of galaxy dynamics and formation, and their own formation and evolution is also a matter of extensive debate. The compact nature of globular clusters means that they are hard to spatially resolve and thus study outside the Local Group. 
In this work we have examined how well EGCs will be detectable in images from the \Euclid telescope, using both simulated pre-launch images and the first early-release observations of the Fornax galaxy cluster. The Euclid Wide Survey will provide high-spatial resolution VIS imaging in the broad \IE\ band as well as near-infrared photometry (\YE, \JE, and \HE). 
We estimate that the 24\,719 known galaxies within 100\,Mpc in the footprint of the \Euclid survey host around 830\,000 EGCs of which about 350\,000 are within the survey's detection limits. For about half of these EGCs, three infrared colours will be available as well. For any galaxy within 50\,Mpc the brighter half of its GC luminosity function will be detectable by the Euclid Wide Survey. The detectability of EGCs is mainly driven by the residual surface brightness of their host galaxy. We find that an automated machine-learning EGC-classification method based on real \Euclid data of the Fornax galaxy cluster provides an efficient method to generate high purity and high completeness GC candidate catalogues. We confirm that EGCs are spatially resolved compared to pure point sources in VIS images of Fornax.
Our analysis of both simulated and first on-sky data show that Euclid will increase the number of GCs accessible with high-resolution imaging substantially compared to previous surveys, and will permit the study of GCs in the outskirts of their hosts. \Euclid is unique in enabling systematic studies of EGCs in a spatially unbiased and homogeneous manner and is primed to improve our understanding of many understudied aspects of GC astrophysics.}

%
%
%
%
   \titlerunning{\Euclid preparation. LVIII. Detecting globular clusters in the Euclid survey}
   \authorrunning{Euclid Collaboration: Voggel et al.}
   
   \maketitle
%
%
%
%
\section{Introduction} 
\label{sec:intro}

Globular Clusters (GCs) exist in almost all galaxies from the most massive ones down to ultra-faint dwarfs. Galaxies can have none or anywhere from less than a handful \citep{Georgiev2010, Karachentsev2013} up to many thousands, as typically found in the brightest galaxies of galaxy clusters \citep[e.g.,][]{Harris1991, Dirsch2003, Strader2006, Harris2013, Durrell2014}. GCs are typically old (ages of the order of $10$\,Gyr) and dense ($r_{\text{h}}<10\,{\text{pc}}$) stellar systems that formed during periods of intense star formation of their current host (or previous host if accreted), and thus we can use them as tracers of this galaxy formation process in the early Universe. 

The upcoming Euclid Wide Survey (EWS) will have high spatial resolution imaging combined with a sky coverage of about 14\,500\,deg$^2$. \Euclid has two instruments and four photometric bands \citep{EuclidSkyOverview,EuclidSkyVIS,EuclidSkyNISP}. The VIS instrument provides a spatial resolution close to \ang{;;0.14} (FWHM), through a single broad red passband \IE\ that covers a range of 5000--9000\,\AA. In addition to the red filter, the near-infrared (NIR) imager NISP covers three bands, \YE, \JE, and \HE \citep{Scaramella-EP1}. In the visible, this new combination of space-based resolution and large survey area allows for the first time the study of a large fraction of the sky at high spatial resolution. This combination enables a first non-spatially biased survey of CGs in the local Universe. The improvement the \Euclid survey will mean for GC science is detailed in this section.

The colour distributions of GC systems are typically bimodal \citep{Peng2006}, yet that difference in colour does not necessarily imply a large age difference, but rather more often a difference in metallicity due to the galaxy enrichment history during their formation \citep{Cantiello2007, Strader2007, Woodley2010}.
While the red GCs in a galaxy follow a radial density distribution similar to the underlying starlight of their host galaxy it has been shown that blue, metal-poor GCs exhibit a more shallow radial drop-off.
The colour of red GCs becomes redder in more luminous (more massive) galaxies while the colour of the blue GCs peak is constant across the range of their host luminosities \citep{Peng2006}. This has been interpreted as evidence that the blue GCs are coming from accreted lower mass galaxies \citep{Cote1998, Pastorello2014}. These differences in colour distributions can trace the ancient accretion history of a galaxy, of which no more short-lived accretion signatures such as tidal tails are remaining. With the wide coverage of \Euclid e.g., differences between red (metal-rich) and blue (metal-poor) GCs as a function of host properties and location in the host galaxy can be studied on a statistically significant sample.

Globular clusters can also be used as tracers of very recent accretion events. In galaxies such as Andromeda \citep{Mackey2013, Mackey2019} or Centaurus\,A \citep{Hughes2021}, GCs have been found to align with the tidal features in the halo. This is direct evidence for the accretion and subsequent deposition of GCs hosted by a smaller galaxy into their new host. Such accreted galaxies may host nuclear star clusters in their centres that are then deposited in the galaxies' halo where they survive typical tidal forces \citep{Bekki2003, Pfeffer2016}. Accreted nuclear clusters have a more extended star-formation history \citep{Kacharov2018} and some also contain central black holes that distinguish them from normal GCs \citep{Seth2014, Voggel2019}. The correlation between nucleus mass and galaxy mass \citep{Georgiev2009} makes remnant nuclei an excellent long-lived tracer of galaxy formation and disruption. Euclid’s depth and coverage with allow a census of a large area of the sky for such accretion events and candidate stripped nuclei.

The appearance of extragalactic GCs as bright compact or point-like sources makes them accessible with spectroscopy out to much larger distances than individual stars. For this reason GCs have been widely used as dynamical tracers of the underlying dark matter halo mass and shape \citep{Frenk1980, vdB1981, Wassermann2018, Reina-Campos2022, Toloba2023}. However, mostly dwarf satellite galaxies are still used as dynamical tracers as they are typically identified to much larger galactocentric radii, whereas GCs are mostly known only in the central 5--10\,kpc of a galaxy. When GCs are rarely identified at radii beyond 50\,kpc, their value as dynamical tracers of underlying dark matter increases significantly because the relative impact of dark matter on GC dynamics increases with radius. Thus once we can detect GCs at larger distances (e.g., \citealt{Veljanoski2014, Dumont2023}), they become a second independent tracer of galaxy dynamics in addition to dwarf satellite galaxies. 

However, detection of reliable GC candidates in the outskirts of local-Universe galaxies has been hindered by either the small fields of view (FoV) of space-based facilities or the low spatial resolution of ground-based facilities, which \Euclid is going to vastly improve upon with its 0.5\,deg$^{2}$ images and stable point-spread function (PSF). Existing ground-based imaging surveys that cover wide areas exist, but they either do not have sufficient depth or they have too high contamination from compact galaxies and foreground stars to provide high-quality GC catalogues. 

Despite the progress, open questions remain. For instance, it is unclear whether the spectral energy distributions (SEDs) of GCs depend on environment, as statistical studies of such potential differences are plagued by spatially varying systematic errors in the heterogeneous existing data  \citep{Powalka2016, Cantiello2018}. 
The stars of GCs in the Milky Way display characteristic variations in their light-element abundances, whose origins have been widely debated, but no consensus has been reached
\citep{Piotto2007,Renzini_etal2015,Gieles_etal2018,Bastian2018}.They produce broader spectral features such as the molecular bands of CN, but also, mostly via associated changes in helium abundance, in the temperature of horizontal branch stars or the relative number of asymptotic giant branch stars \citep{Sandage1967, DAntona_etal2002,Lagioia2021}. To detect the resulting small differences in the light of GCs beyond the Local Group, homogeneous and precise photometry is needed like \Euclid will be able to provide it with its stable PSF. A detailed \Euclid photometric catalogue of nearby GCs will also help improve the ingredients of stellar population synthesis models, the SED-predictions of which differ by amounts similar to current systematic differences between present-day heterogeneous surveys, in particular when near-infrared wavelengths are also considered \citep{Powalka2016, Verro2022}.

For systems beyond the Local Group, the biggest challenge is to separate the GCs from contaminants, which can be foreground stars or compact background galaxies. For distances beyond about 5\,Mpc, GCs appear like point sources for typical ground-based imaging unassisted by adaptive optics. Even in space-based images their light profile can only be resolved out to host distances of 30--40\,Mpc. Therefore, multiple colours or follow-up spectroscopy are needed to confirm whether a source is a GC or not.

The main limitation for spectroscopy and space-based imaging is observation time. Typically for a nearby galaxy that spans several degrees on the sky, deep photometry catalogues can contain  millions of sources. Deriving the spectroscopy for such large numbers of objects, including the faintest in the area, is presently infeasible with existing observing facilities. When using colours, the combination of near-ultraviolet, optical, and near-infrared data is efficient, but expensive \citep{Munoz2014}. Fortunately, even minimal information on source compactness can reduce the need for multiple photometric passbands \citep{Voggel2020,Buzzo2022, Harris2023}. The \Euclid spatial resolution is therefore a key ingredient to reach a pure and volume-complete census of GCs.
The downside of past space-based studies such as those pioneered by the \textit{Hubble} Space Telescope is the limited FoV of a few arcminutes. That led to `postage stamp' studies, which rarely extended out beyond two or three effective galaxy radii in the local Universe. The results were limited data sets of the central areas of galaxies that could not be easily compared to wider ground-based data sets of GC properties. With \Euclid’s coverage of one-third of the sky in a spatially unbiased manner the detection of intracluster GCs as a function of cluster and GC properties will become possible.

\section{Number of expected globular clusters for galaxies in the EWS footprint} \label{sec:numbers}

The goal of this section is to derive an order of magnitude estimate of the number of GCs detectable in \Euclid imaging. We will do this by using well-established GC number scaling relations to calculate the theoretical numbers of GCs and their magnitudes in the known galaxies in the local Universe, out to a maximum distance that we set to 100\,Mpc.

The globular cluster luminosity function (GCLF) peaks at $M_V=-7.5$\,mag and is typically Gaussian \citep{Rejkuba2012}, which roughly corresponds to an absolute AB magnitude of $M_{\IE}=-8$\,mag in the \IE\ passband of the \Euclid VIS instrument. The transformation is based on integrated GC spectra computed with population synthesis models \citep{Fioc1997, Verro2022}, for which synthetic photometry predicts that $(V-\IE)$ [AB] varies between 0.3 and 0.6 at ages between 8 and 13\,Gyr, depending on metallicity. 

At 100\,Mpc, the brightest GCs with $M_{\IE}=-10.5$ will have an apparent AB magnitude of $\IE\simeq 25$ which is just one magnitude brighter than the $5\,\sigma$ point source detection limit of $\IE=26.2$ in the EWS \citep{Scaramella-EP1}. 

We exclude galaxies within the Local Group as their globular clusters will be resolved into many individual stars and thus many separate sources in the \Euclid catalogue, or have complex morphologies. This partial resolution of GCs is expected out to around 3\,Mpc. We focus on GCs that will appear like point-sources at distances between 3 and 100\,Mpc but are slightly spatially resolved when looking at them in more detail.

\begin{figure*}
   \centering
   \includegraphics[width=\hsize]{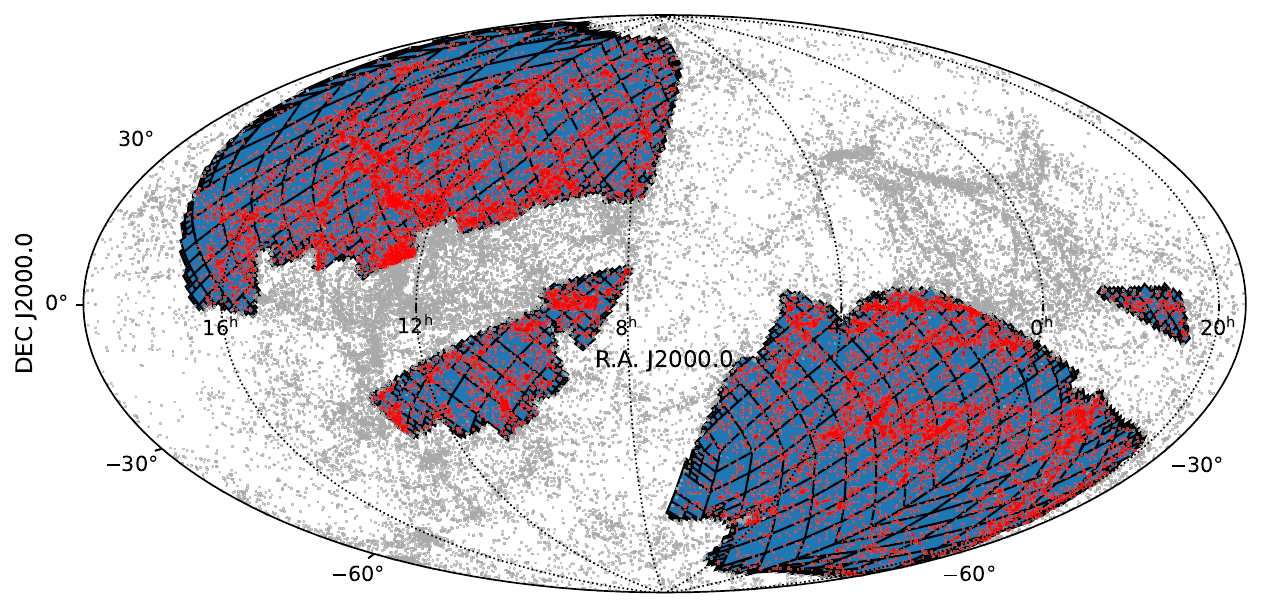}
     \caption{\Euclid full 6-year survey area is shown as a \texttt{HEALPix} grid in blue. The galaxies in the HECATE catalogue within 100\,Mpc distance are shown as points, those within the footprint in red, and those outside in grey.} 
         \label{fig:footprint}
   \end{figure*}

As a first step, we thus need to identify how many local galaxies out to 100\,Mpc are expected to be observed in the footprint of the planned EWS. We select all galaxies from  the Heraklion Extragalactic Catalogue of local-Universe galaxies \citep[HECATE,][]{Kovlakas2021} that fall within the EWS region. We find a total of 24\,719 galaxies that have a distance measurement and are  within a distance of 100\,Mpc, and have a listed $B$ magnitude (see Fig.~\ref{fig:footprint}). We can only use galaxies with a distance measurement because their absolute magnitude is required to make a prediction of how many GCs a galaxy is expected to host. GCs can be confirmed much easier in galaxies with a known distance, so this selection assures we actually predict how many GCs are detectable in the confirmed local Universe galaxies in the EWS.

\begin{figure}
   \centering
   \includegraphics[width=\hsize]{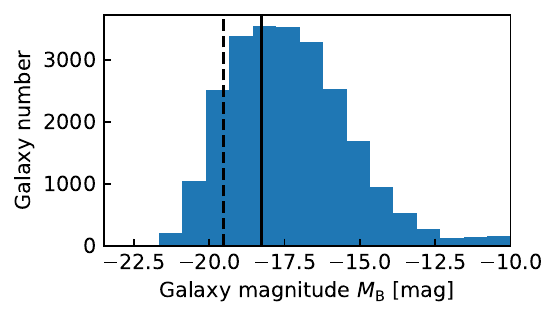}
     \caption{Distribution of absolute $B$ magnitudes of HECATE galaxies within the planned \Euclid footprint that are also within 100\,Mpc. The 100\% magnitude completeness limit out to 33\,Mpc (solid) and to 100\,Mpc (dashed) are shown as vertical lines. } 
         \label{fig:maggals}
   \end{figure}

We use \texttt{MOCPy} \citep{mocpy} to filter the list of known galaxies with the complex multi-object coverage map (MOC) of the \Euclid full 6-year survey area. The distribution of the absolute $B$ magnitudes of all selected galaxies is shown in Fig.~\ref{fig:maggals}. The HECATE galaxy catalogue is complete for galaxies brighter than $M_B=-18.27$  within 33\,Mpc, and out to 100\,Mpc the completeness limit is $M_B=-19.52$ \citep[see][]{Kovlakas2021}. Thus, we are complete at the bright end of the galaxy distribution to roughly the mass of the Milky Way galaxy, which is the mass range where the vast majority of the total number of GCs is expected to reside \citep{Harris2016}.

\subsection{Specific GC frequency}   
 To estimate how many GCs we expect in all local galaxies selected above, we model their expected number of GCs by using what we know about the specific frequency of GCs in galaxies from empirically determined relations. The number of GCs in a given galaxy relates to its total luminosity and this relation is commonly expressed with a specific frequency, 
 \begin{equation}
 S_{\!\sfont{N}} = N_{\sfont{GC}} \; 10^{0.4\,(M_V+15)}\;,    
 \end{equation}
 where $N_{\sfont{GC}}$ is the number of GCs and $M_V$ the total magnitude of the host galaxy \citep{HarrisVdBergh1981}. With an assessment of $S_{\!\sfont{N}}$ as a function of host galaxy magnitude we can reverse this relation to predict GC numbers.

First, we assemble a catalogue of galaxies for which the total number of GCs in their system is known. The starting point for this is the large compilation of GC numbers in 422 galaxies in the local Universe \citep{Harris2013}, with updated total GC numbers for Centaurus\,A \citep{Hughes2021} and M32 \citep{Karachentsev2013}. We add well-studied dwarf galaxies to the sample \citep{Georgiev2009, Karachentsev2015, Crnojevic2016, Cole2017, Caldwell2017}. We also add data from recent studies of GCs in ultra-diffuse galaxies \citep{vanDokkum2017, Lim2018, Amorisco2018, Muller2021, Saif2022}.

      \begin{figure}
   \centering
   \includegraphics[width=\hsize]{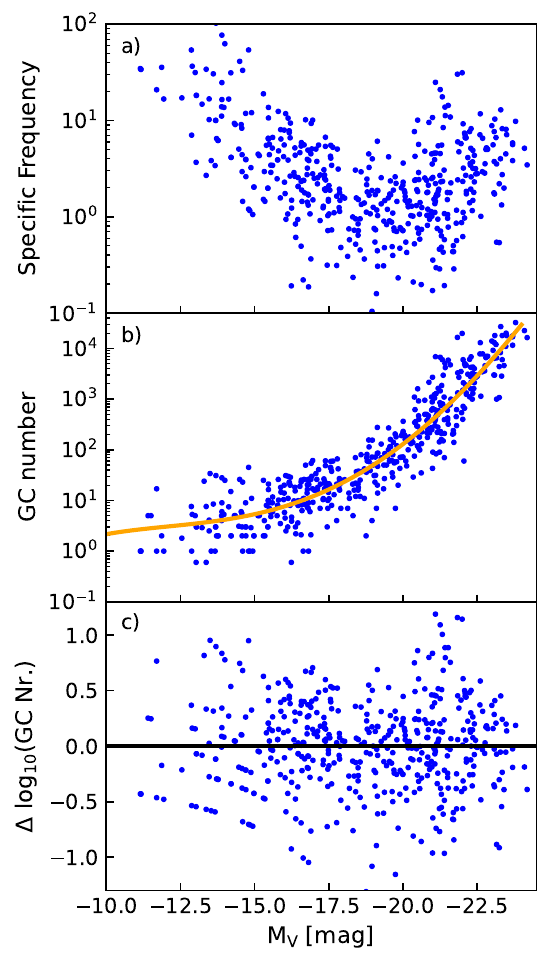}
     \caption{Predicted GC numbers as a function of magnitude. \textit{Panel a):} Specific frequency of GCs as a function of host galaxy magnitude for our galaxy catalogue. \textit{Panel b): } Total GC number in each galaxy as a function of its magnitude. A polynomial fit to the data is shown as orange line. \textit{Panel c): } Log difference between the mean GC number for a given galaxy magnitude (orange fit in the  middle panel) and the actual GC number of each galaxy. The alignment of data points in the bottom half is a natural consequence of the small numbers of GCs where the difference is only a single or two GCs for those low-luminosity galaxies.  } 
         \label{fig:specfreq}
   \end{figure}

      \begin{figure}
   \centering
   \includegraphics[width=\hsize]{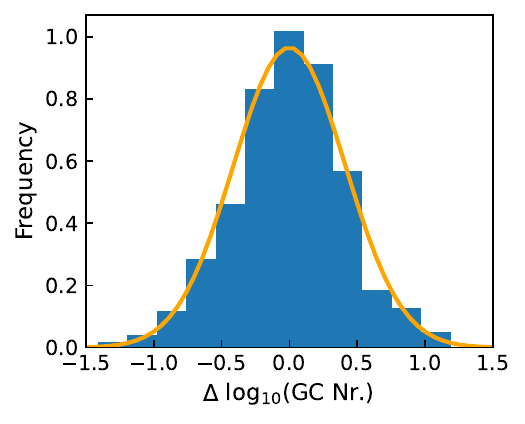}
     \caption{Distribution of the difference between log of the measured GC number in a given galaxy and their expected mean log(GC) number from the fitted relation (see orange fit in Fig.~\ref{fig:specfreq}).  The distribution is log normal with a mean of 0 and a standard deviation of $\sigma=0.41$.} 
         \label{fig:logdiff}
   \end{figure}  

In Fig.~\ref{fig:specfreq} we show $S_{\!\sfont{N}}$ in the compiled catalogue as a function of galaxy luminosity in the top panel and the total number of GCs as a function of magnitude in the middle panel. We represent the variations of the average number of GCs with galaxy luminosity with a 4th-order polynomial fit (orange line). The residuals in the bottom panel are clearly distributed in a log-normal form, as shown in Fig.\,\ref{fig:logdiff} and has a mean of 0 and a standard deviation of $\sigma=0.41$\,dex. This simple statistical behaviour of $S_{\!\sfont{N}}$ is used in the next step to anticipate GC numbers.

\subsection{Number of observable GCs in \Euclid}

To predict the total number of GCs expected in the EWS footprint, we randomly draw a total number of GCs for each galaxy in our HECATE sample, based on its magnitude and on the $S_{\!\sfont{N}}$ distribution just described. The drawn Nr of GCs is then rounded to the nearest integer. In the case of the faintest galaxies this expected average GC number can be very small ($<5$) and including the log-normal scatter in GC system size, it can predict $<0.5$ GCs and thus the rounded number will be zero. The existence of dwarf galaxies that host zero GCs is naturally included in this model. The fraction of low-mass galaxies hosting no GCs is not understood observationally at this point. We do not have the statistics to deduce how common they are as we lack deep imaging needed to cover the whole GC luminosity function in a large set of dwarf galaxies at the moment. We note that with Euclid we expect progress in understanding the GC luminosity function (or the absence) in dwarfs and first observational results have been shown in \citep{EROFornaxGCs}. As the number of GCs in these very faint galaxies is negligible compared to the total expected number the effect of this approximation is very small and within the uncertainties of the final forecast number.

\begin{figure}
   \centering
   \includegraphics[width=\hsize]{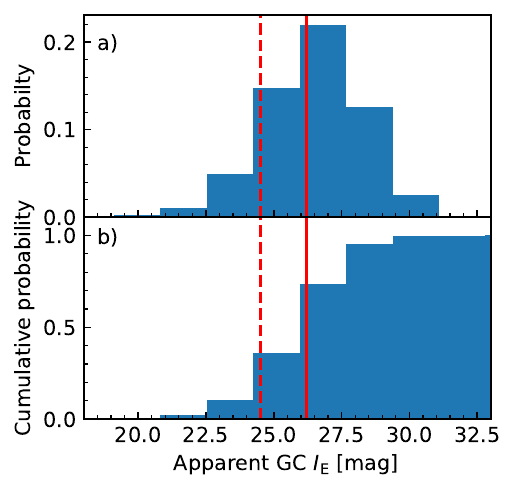}
     \caption{Magnitude distribution of all predicted GCs. \textit{Panel a):} Distribution of apparent magnitudes of all expected GCs hosted by galaxies out to 100\,Mpc. \textit{Panel b):} Cumulative distribution of the relative number of GCs. The solid red line marks the VIS point source detection threshold of 26.2 and the dashed line marks 24.5 as the magnitude limit in which GCs will also be detected in NISP (see Fig.~\ref{fig:SED_and_sensitivity}).} 
         \label{fig:mag_dist_gcs}
   \end{figure}

We then randomly distribute this number of GCs among the typical magnitude distribution of GC systems, i.e., a normal distribution with a mean absolute magnitude of $M_V=-7.5$ and a standard deviation of $\sigma=1.2$\,mag \citep{Rejkuba2012}. The width of this GC luminosity function depends on the galaxy luminosity and is wider for brighter galaxies \citep{Villegas2010}, yet for our estimate of GC numbers that difference is negligible as other uncertainties such as the galaxy incompleteness at the faint end dominate. For each galaxy we now have a representation of the absolute magnitudes of each GC they are expected to host, and we use the distance information from the HECATE catalogue to predict how they would appear in VIS images. Figure~\ref{fig:mag_dist_gcs} displays the distribution of apparent magnitudes of all expected GCs in the EWS footprint, for one such stochastic simulation. 

The GC detection threshold of $\IE=26.2\,{\rm mag}$ in Fig.\,\ref{fig:mag_dist_gcs} shows that close to half of all GCs in local-Universe galaxies are detectable with \Euclid. We ran the stochastic simulation pipeline 100 times and we took the average of the predicted total number of GCs among the runs and the standard deviation of all realisations. We expect an average of 
$830\,000 \pm 12\,000$ GCs in the known galaxies in the \Euclid footprint, of which 
$350\,000 \pm 6\,000$ ($44\%$) are above the formal detection threshold and $115\,000\pm1500$ will have the excellent photometric quality that comes with $\IE\leq 24.5$.
For example, a typical Milky Way-like galaxy at $M_{V}=-21$ is expected to host 500 GCs, of which we will detect 275 (55\%) if the galaxy lies at 50\,Mpc from us and we will still detect 100 GCs at 100\,Mpc.

The errors quoted here are statistical and take into account the scatter in the total number of GCs at a given galaxy luminosity. There is systemic uncertainty in the derived GC numbers due to incompleteness in our galaxy luminosity function, host galaxy morphological type, and environment. The total GC number can depend on all these factors. Contamination by galaxy light, especially close to the galaxy centre will lower our ability to detect them so the realistic number that is detectable is likely lower.

\begin{figure}
   \centering
   \includegraphics[width=\hsize]{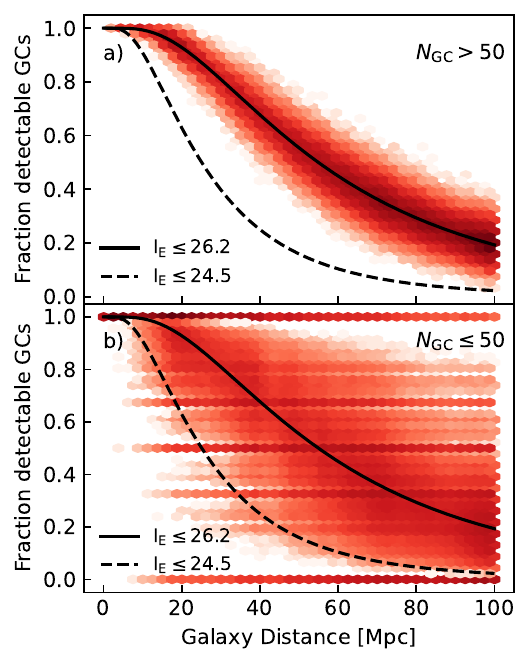}
     \caption{Fraction of simulated GCs that are above the detection threshold as a function of distance, for each galaxy in the HECATE-based sample. The two-dimensional density represented with red shaded bins summarizes the theoretical GC detectability fraction of all galaxies in 100 Monte-Carlo realisations (i.e., 100 $\times$ 26\,711 galaxies). The solid line is the analytic predicted fraction for $\IE\leq26.2$ and the dashed line is for $\IE\leq25$. \textit{Panel a):} Only systems with $N_{\sfont{GC}}>50$. \textit{Panel b):} Only GC systems with $N_{\sfont{GC}}\leq50$ to illustrate the stochasticity introduced by small galaxies and their small total number of GCs.} 
         \label{fig:frac_gcs}
\end{figure}  
   
To get an idea of how the theoretical detectability varies for individual galaxies with distance, we show in Fig.\,\ref{fig:frac_gcs} the fraction of a given GC system for individual galaxies that is detectable adopting the limit of $\IE=26.2\,{\rm mag}$. Both panels in Fig.~\ref{fig:frac_gcs} show 100 Monte Carlo simulations of the whole set of local-Universe galaxies as shaded bins. From the drawn GC luminosity distribution we predict those as detectable that are brighter than the threshold. The solid and dotted lines are the analytic predicted fractions that are detectable at a given distance for $\IE=26.2$ and $25$, respectively. Only systems with $N_{\sfont{GC}}>50$ are shown in the top panel and galaxies with small GC systems of $N_{\sfont{GC}}\leq50$ are shown in the bottom panel. The bottom panel's large spread is the natural consequence of small-number statistics, and in particular the clustering at fractions of 1/3, 1/2, and 2/3 are owed to that.

As expected, the fraction of GCs that are brighter than $M_V=26.2$ without considering confusion with the host galaxy decreases with distance. In terms of limiting brightness, \Euclid is capable of examining almost the entire GC system of a galaxy within 20\,Mpc. For galaxies at 50\,Mpc, the detectability drops to 50\% meaning that only the bright half of the GC luminosity function is accessible. The typical uncertainty in the detectability for large GC systems is around $10\%$.
Even at 100\,Mpc, \Euclid can still detect the roughly 20\% brightest GGs. Beyond 100\,Mpc we thus only expect ultra-compact dwarf-like objects that are brighter than $M_V=-10$ to be present in the VIS images of the EWS.

\subsection{Availability of infrared colours in the \Euclid survey}
\label{sec:NISP_avail}

      \begin{figure}
   \centering
   \includegraphics[width=\hsize]
   {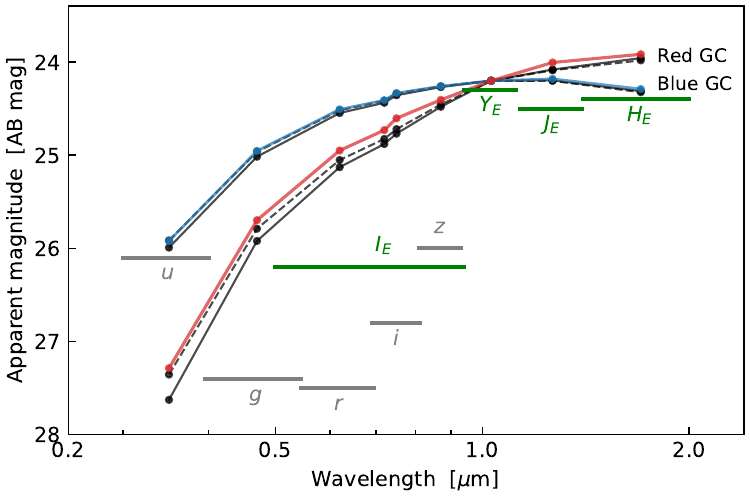}
     \caption{Spectral energy distribution of a typical red GC ($[M/H] \simeq -0.2$) and blue GC ($[M/H]\simeq -2$), as predicted at ages of 8\,Gyr (dashed) or 13\,Gyr (solid) with two population synthesis models. In red, the detection limits of the four \Euclid passbands \IE, \YE, \JE, and \HE\ are shown, together with the 10\,yr LSST survey limits  for the $ugriz$ bands.}
         \label{fig:SED_and_sensitivity}
   \end{figure} 
   
The optical to NIR spectral energy distributions of typical red and blue GCs, as predicted with two population synthesis codes, are displayed in Fig.~\ref{fig:SED_and_sensitivity}. The black curves are obtained with \texttt{P\'egase} \citep[][low spectral resolution version]{LeborgnePeg2004,PegaseCode2011}, and the red and blue curves with the combination of \texttt{Parsec} isochrones and X-shooter spectra of \citet{Verro2022}. In green the $5\,\sigma$ point-source detection limits of the EWS are shown \citep{Scaramella-EP1}, together with the LSST 10-year sensitivity limits in $u,g,r,i$, and $z$ \citep{LSST2019}, in grey. The GC curves are normalised to $\YE=24.2$, which puts the model GC above the detection threshold in the three NIR bands. 

It follows that GCs with a magnitude of at least $\IE\leq24.5\,{\rm mag}$ will also be detected in the three \Euclid NIR bands. Using the simulations of the previous section, we find $120\,000\pm 3000$ GCs pass this threshold. This represents about 34\,\% of the 353\,000 GCs with expected detections in VIS images.

Based on signal-to-noise curves that indicate PSF-photometry errors of 0.2\,mag at $\IE=26.2\,{\rm mag}$ \citep{Scaramella-EP1}, \Euclid VIS photometric errors will be smaller than 0.05\,mag for quasi-point sources brighter than 24.5. With this uncertainty on \IE, random errors on colours that combine an \IE\ magnitude with a magnitude from a NISP passband will be smaller than 10\,\% when NISP errors are smaller than about 0.09\,mag; signal-to-noise curves that predict $5\,\sigma$ detections at 24.4\,mag \citep{Scaramella-EP1} indicate that this occurs for NISP magnitudes brighter than about 23.4\,mag. This is about 1.5\,mag deeper than specialised wide-area ground-based surveys such as NGVS-IR \citep{Munoz2014}. 

The key role played by the EWS will be to provide uniformity over vast areas of the sky, allowing past and future pointed  observations from  either the ground or space to be calibrated to a common scale. 

Figure~\ref{fig:SED_and_sensitivity} shows that all the Southern Hemisphere \Euclid extragalactic GCs detected in NISP images will progressively benefit from excellent LSST $g,r,i$, and $z$ measurements, but only the bluer ones measured precisely by \Euclid will be detected in the $u$ band. In the Northern Hemisphere, the only ongoing very wide $u$-band survey is UNIONS-$u$ \citep{Ibata2017}, the 10-$\sigma$ depth of which is about 23.5\,mag. Therefore in the Northern Hemisphere large-area EGC studies will have to rely exclusively on the \Euclid data themselves.

\section{Simulated VIS images with artificial globular clusters}
\label{sec:imgs}
\subsection{Simulation method}
In order to better understand the number of GCs that will be actually detectable, we used simulated \Euclid VIS images \citep{Serrano2024} and we inserted local-Universe galaxies and GCs in them. We then ran those images through the Euclid Consortium pipeline processing functions that combine four VIS exposures to create a mosaic, subtract extended background emission, and produce catalogues. The pipeline's individual mosaic images are smaller than a VIS field of view (they are square with a side of $32\arcminute$) and have pixels of $\ang{;;0.1}$ size.

Our added objects are positioned in individual VIS exposures such that they fit within one mosaic. We simulated four configurations (labelled {\tt FoV1} to {\tt FoV4}), in which we added galaxies and their corresponding GCs at several different distances: 5\,Mpc, 10\,Mpc, and 20\,Mpc for {\tt FoV1} to {\tt FoV3}, and a combination of distances between 40\,Mpc and 60\,Mpc in {\tt FoV4} (with a few exceptions at 20\,Mpc, as specified below).

The injection of globular clusters and galaxies into simulated VIS exposures is done through stamps generated for a given RA and DEC location on the different CCDs by the \texttt{GalSim} Python package \citep{Galsim2015}. It enables the convolution of injected objects by the VIS PSF and the addition of Poisson noise to these stamps. The native \texttt{GalSim} package contains a S\'ersic function used for galaxies detailed later. For the EGCs, we have written a custom function to allow \texttt{GalSim} to use the King model \citep{King1962}.

All added galaxies were given a S\'ersic profile with an index of $n=2$, an ellipticity of 0.3, an effective radius of $r_{\rm eff}=1$\,kpc, and an absolute magnitude of $M_{\IE}=-18.5$\,mag, but projected to the required distance. In {\tt FoV4}, 29 of the standard galaxies were replaced with modelled ultra-diffuse galaxies, with $M_{\IE}=-14$ and $r_{\rm eff}=2$\,kpc, placed at 
a distance of 20\,Mpc. 
As the apparent sizes of the galaxies shrink with distance, the simulations of further distances include more galaxies overall.

On top of these galaxies we add GCs with varying magnitudes and colours, all represented with King models. Their \IE\  magnitudes are chosen such that they cover the GC luminosity function in steps of 0.5\,mag, within the range detectable in \Euclid VIS images at a given distance. The range of absolute and apparent magnitudes for each distance are listed in Table~\ref{tab:simulation}. 

At magnitudes brighter than $\IE=18.5$ sources start to saturate, which is a concern in {\tt FoV1} at 5\,Mpc where any GC brighter than $M_{\IE} =-10$ would fall into this range. Therefore, luminosities of simulated GCs in {\tt FoV1} are truncated to $M_{\IE}\leq -9$\,mag, whereas for all other distances the brightest GC is between $M_{\IE}=-10$ and $M_{\IE}=-11.5$ depending on distance. This covers the magnitude range up to objects with magnitudes brighter than $M_V=-10$ that are similar to ultra-compact dwarfs (UCDs). Our faintest added GCs have $\IE = 25.5$, which translates to $M_{\IE}=-6.0$ and $M_{\IE}=-8.5$\,mag, respectively, in {\tt FoV3} and {\tt FoV4}. 

\begin{figure}
\centering
\includegraphics[width=\hsize]{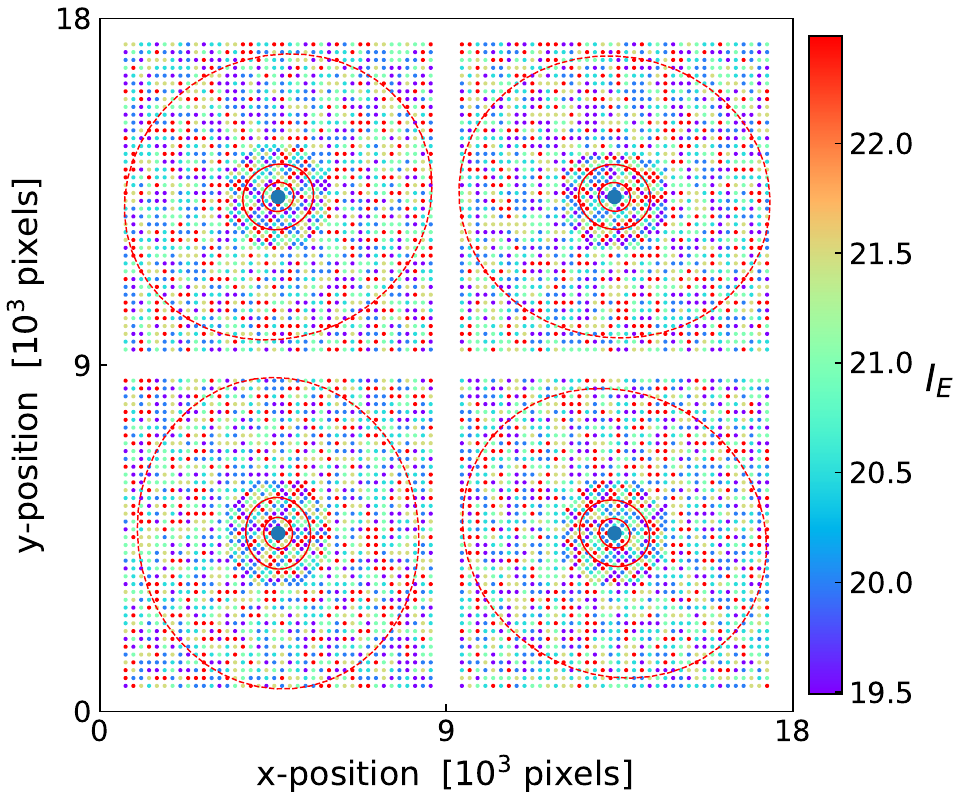}
\caption{Location of artificial galaxies and GCs that were added to simulated \Euclid pointings, which already contained foreground stars and background galaxies. The host galaxies are represented by surface-brightness contours marking the half-light radius, $\mu=23\,\rm{mag}\,\rm{arcsec}^{-2}$, and $\mu=29.5\,\rm{mag}\,\rm{arcsec}^{-2}$. The artificial GCs are distributed in a grid pattern, coloured according to their magnitudes. This image is {\tt FoV1} for which the distance adopted for the added objects is 5\,Mpc. The coordinates are shown in units of $\ang{;;0.1}$ pixels and the colour bar is the apparent \IE\ magnitude of the injected GCs.} 
\label{fig:fov1}
\end{figure}

We adopted a typical GC core radius of $r_{\rm c}$=1.2\,pc and a King concentration index of $r_{\rm t}/r_{\rm c}=1.4$, where $r_{\rm t}$ is the tidal radius where the density falls to zero for GCs \citep{McLaughlin2005}. In all simulations except the one at 5\,Mpc one, we also included larger size GCs with core radii of 4\,pc. The host galaxy properties and the GC ranges are summarised in Table~\ref{tab:simulation} and the arrangement for {\tt FoV1} is shown as an example in Fig.~\ref{fig:fov1}.

\begin{table*}
\caption{Summary of the simulated galaxies and GCs.}
\begin{center}
\smallskip
\label{tab:simulation}
\smallskip
\begin{tabular}{ccccccc}
  \hline\hline
  \noalign{\vskip 1pt}
FoV & Distance [Mpc] & Galaxy & $N_{\text{gal}}$ & $N_{\sfont{GC}}$ & Range of $M_{\IE}$ for GC [mag] & Range of $\IE$ for GC [mag]  \\
  \hline
  \noalign{\vskip 1pt}
{\tt FoV1} & 5 & Normal & 4  & 6972 & $-9.0$ \dots $-4.5$ & 19.5 \dots 22.5  \\
{\tt FoV2} & 10 & Normal& 16 & 5640 & $-10.0$ \dots $-6.0$ & 20.0 \dots 24.0 \\
{\tt FoV3} & 20 & Normal & 64 & 6960 & $-11.5$ \dots $-6.0$ & 20.0 \dots 25.5  \\
{\tt FoV4} & 40 & Normal & 111 & 2800 & $-11.5$ \dots $-7.5$ & 21.5 \dots 25.5 \\
{\tt FoV4} & 60 & Normal & 115 & 1035 & $-11.4$ \dots $-8.4$ & 22.5 \dots 25.5 \\
{\tt FoV4} & 20.5 & UDG & 29 & 750 & $-10.0$ \dots $-7.0$ & 21.5 \dots 24.5  \\
\hline
\end{tabular}
\footnotesize
 \tablefoot{Column~1 recalls the labels of our four simulations. Column~2 is the distance at which the added galaxies were placed and Col.~3 specifies whether these galaxies are standard (see text) or the UDG model. Column~4 gives the number of host galaxies at this distance, and the Col.~5 the total number of GCs distributed across all host galaxies. Columns~6 and 7 are the magnitude range of added GCs in absolute and apparent \IE\ magnitudes, respectively.}
\end{center}
\end{table*}

\subsection{Globular cluster detection in simulated VIS images}

The most straightforward way of identifying GCs is not to analyze the vast data set of VIS images that will be taken by \Euclid, but rather to use the data catalogue that will be created by the \Euclid analysis pipeline automatically. This pipeline-catalogue contains the merged information for all extracted sources from both VIS and NISP. As such, it contains fluxes in the \IE, \YE, \JE, and \HE\ bands, positions, morphological information such as their compactness, central surface brightness, or their likelihood to be a star or a galaxy. In this catalogue a large quantity of information on the detected sources is already present. This pipeline\footnote{We use the version available in 2023.} was also run on our simulated VIS data after the addition of the galaxies and GCs.

Using these output catalogues for artificial GCs will give us an idea of how such objects compare to other sources in the field, and thus what purity and completeness we can expect of a GC candidate catalogue derived from this source catalogue. In addition this gives us the opportunity to compare how accurately the pipeline recovers the known input properties of the GCs. 

To this aim we match the input catalogue of artificial-GC positions with the output pipeline-catalogue. We use a small matching radius of $\ang{;;0.1}$ 
to avoid random field contamination. The average distance of our matched positions is $\ang{;;0.05}$, corresponding to half a pixel in VIS and thus indicating that we indeed retrieved the input GC sources. 
Overall we find that the automatic \Euclid pipeline was able to detect 18\,340 GCs compared to the 24\,154 artificial GCs that were injected, for an overall recovery fraction of 76\%.

The fact that these GCs get recovered by the pipeline and included in the catalogue does not necessarily imply that we can identify them as bona fide GCs right away. For this they will have to be studied more carefully for their light profile or followed up by spectroscopy. However, the recovery fractions indicate how many we can expect in the catalogue to begin with, based on which we generate a GC candidate list. Those GCs that are not directly recovered by the Euclid photometric pipeline will still exist in the images if they are brighter than the detection threshold. The large spatial coverage of the \Euclid survey means that detecting them in a systematic manner will be challenging because it requires treating each image separately and for example removing the light of the surrounding galaxy.

\subsection{Recovery fraction as a function of local surface brightness and GC magnitude}

\begin{figure*}
\sidecaption
\includegraphics[width=12cm]{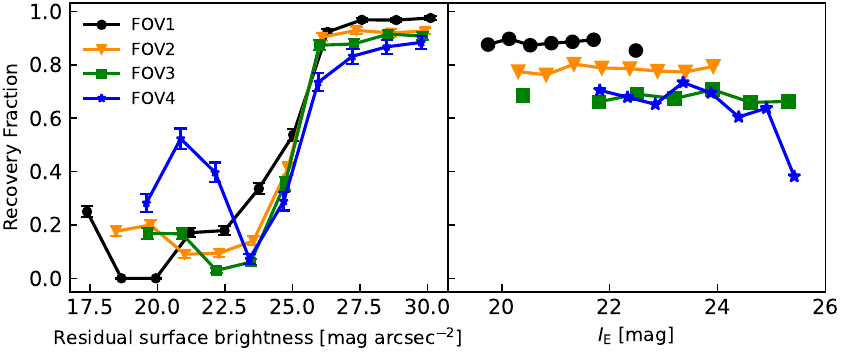}
\caption{Recovery fraction of GCs. \textbf{Left panel:} Fraction of added GCs that have been detected by the \Euclid photometric pipeline as a function of the local residual surface brightness at the location of the GC. The four different symbols are denoting the different configurations of the simulations. \textbf{Right panel:} Fraction of detected GCs by the \Euclid photometric pipeline as a function of the total input GC $\IE$ magnitude. The Color codding is the same in both panels.}  
\label{fig:radial}
\end{figure*}

The left panel of Fig.~\ref{fig:radial} shows how the GC detectability depends on the local surface brightness at the GC location and the input \IE\ magnitude of the GCs. The local surface brightness at the position of each inserted GC is derived by calculating the surface brightness in an annulus between $r=\ang{;;2}$ and $r=\ang{;;3}$. The error bars shown are calculated as the 95\% binomial Wilson confidence intervals.

The \Euclid pipeline automatically removes foreground galaxy light if it spans more than one CCD. This is the case for {\tt FoV1} and {\tt FoV2} and thus here the calculated surface brightness is the background of the images plus the residual galaxy surface brightness from this removal. In the two most distant simulations, the galaxies are smaller and the pipeline does not attempt to remove the light and thus the calculated surface brightness includes the background plus the galaxy.

The fraction of recovered GCs is mostly below 20\% for a surface brightness brighter than around $25\,\text{mag}\,\text{arcsec}^{-2}$ and then increases sharply to an 80--90\% recovery fraction when the local surface brightness drops below that. This surface brightness threshold is the same for each simulation and thus independent of distance. For the most distant simulation {\tt FoV4} we find a spike for the recovery fraction at $21\,\text{mag}\,\text{arcsec}^{-2}$, which is likely due to the detection of the inserted nuclear clusters in these simulations. 

The surface-brightness of $25\,\text{mag}\,\text{arcsec}^{-2}$ at which the detectability is very high corresponds to about 4 times the effective radius of the simulated galaxy in {\tt FoV3} and {\tt FoV4}, where the pipeline does not automatically subtract the galaxy light. In the two closest simulations {\tt FoV1} and {\tt FoV2} this limit is closer to the core of the galaxy as the pipeline removes galaxy light from foreground galaxies that span at least an entire detector.

We conclude that host galaxy light subtraction is required for the detection of GCs close the the host centre. For nearby ($<20$\,Mpc) galaxies, the \Euclid pipeline does this in an automated, albeit not perfect manner, whereas for galaxies at larger distances this will have to be done separately. Only for ultra-faint dwarfs is this effect negligible, as their central surface brightness levels are not much larger than $25\,\text{mag}\,\text{arcsec}^{-2}$.

We also tested whether there was a general magnitude dependence of the recovery fraction. The overall detectability is shown in the right panel of Fig. \ref{fig:radial}, where we find that the recovery fraction does not depend directly on GC magnitude when they are brighter than $\IE=25$. Only when the GC magnitude is close to the VIS detection threshold we observe a sharp drop-off in the detectability of a given GC; in our simulations this is the case for our faintest GC magnitude bin, at $\IE=25.5$. At all GC magnitudes brighter than this threshold, the detectability is mainly driven by the location of the GCs in terms of local surface brightness, which in the sky results from a combination of distance to the host galaxy and galaxy light profile. This emphasizes the benefits one may obtain from galaxy subtraction prior to detection, a feature that is not implemented in the standard \Euclid pipeline because it is not a requirement of its core science program. The majority of GCs that we will detect with the standard pipeline will be in the outskirts of the galaxies and not within their half-light radius. 

Finally, it is worth noticing that in {\tt FoV4} the recovery fraction for the GCs placed on low surface-brightness galaxies is 96\%. \Euclid will excel at detecting the GCs of dwarf galaxies. We refer to  \cite{EROFornaxGCs} and \cite{EROPerseusDGs} for first results in this area, based on early observations of both the Fornax and the Perseus galaxy clusters.

\subsection{Magnitude accuracy of the recovered globular clusters}

\begin{figure}
\centering
\includegraphics[width=\hsize]{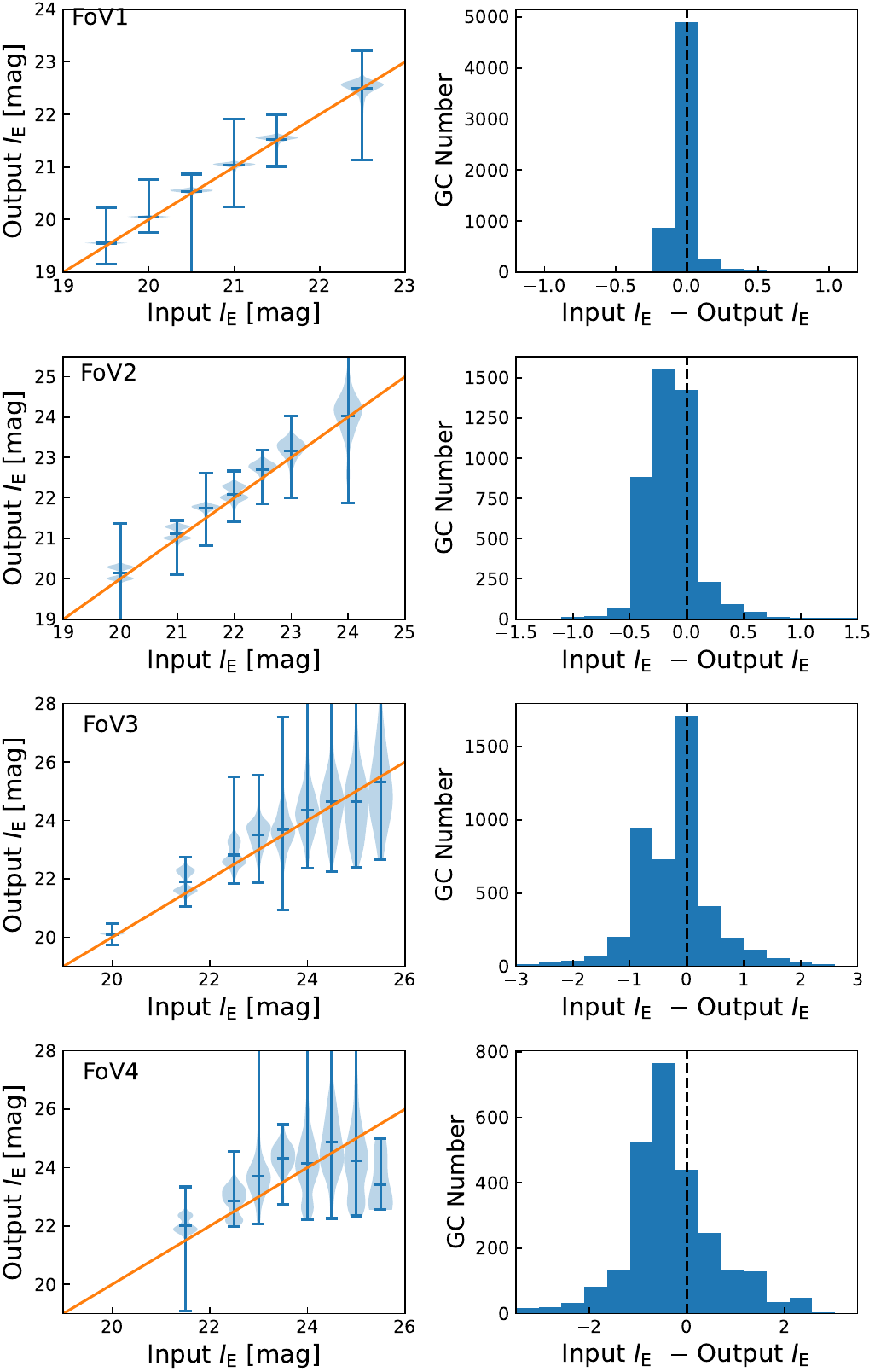}
\caption{Comparison of magnitudes of simulated GCs. \textit{Left column:} Comparison of input and output \IE\ magnitudes of the simulated GCs. The distributions of measured magnitudes from the \Euclid photometry pipeline are plotted against the input values, with the mean and the extremes marked with horizontal bars. From top to bottom we show the results for simulated {\tt FoV} 1, 2, 3, and 4, which correspond to increasing galaxy distance. The solid line marks the 1:1 relation. \textit{Right column:} Histograms of the differences between the input and output $\IE$ magnitudes for the GCs for the same four {\tt FoV}.}
\label{fig:mag_accuracy}
\end{figure}

Here we provide a basic test of how well the input GC properties are recovered by the automatic \Euclid pipeline and how accurately the recovered magnitude compares to the input magnitude. We show the comparison of input VIS magnitude against the recovered VIS aperture photometry in the left column of Fig.~\ref{fig:mag_accuracy}. The aperture diameter used for the ouput VIS photometry is twice the FWHM of the PSF. We use aperture photometry instead of PSF photometry as our mock GCs are substantially more extended than the PSF. This figure shows only those objects that had a match in the output data and does not account for those that were not detected by the photometric pipeline. Clusters inserted exactly at the centre of a galaxy were also removed, as their recovered magnitudes are inevitably too bright.

As expected, the precision of the retrieved magnitudes decreases with increasing distance of the modeled GC system. The standard deviation of the retrieved magnitudes is around 0.1\,mag for the closest {\tt FoV1} at 5\,Mpc but increases to 1.8\,mag for {\tt FoV4} that covers distances of 40 to 60\,Mpc.
Out to 20\,Mpc distance the mean retrieved magnitude is in very good agreement with the simulated input values. The mean magnitude becomes brighter than the input value for the faintest GCs in the two more distant simulations due to injected GCs approaching the detection threshold at 26\,mag. Thus, close to the detection threshold the retrieved magnitudes are likely biased due to contamination by noise and neighbouring objects.

An important difference between the nearby simulations at 5 and 10\,Mpc and the more remote ones, is that the nearby host galaxies span more than one VIS-detector chip, and some of the emission in their outer parts is removed by the background-subtraction algorithm of the default pipeline. For the two more distant simulations the size of the simulated host galaxies is smaller, so that their light remains present in the mosaics. This is likely a major source of added uncertainty in the retrieval of GC magnitudes for the two simulations at larger distances. 

In this section, we use the simulated VIS mosaics exactly as they are returned by the pipeline. We could subtract the galaxy light for the smaller galaxies, yet this type of manual intervention into the imaging data will require a very large computational cost due to the vast data size of the full EWS. Thus we want to validate our GC detection accuracy on standard \Euclid data products. 
However, in the future it might be useful to remove the host galaxy light in an additional step for selected sections of the full \Euclid survey that are of particular interest to EGCs, such as in galaxy clusters.

\subsection{Maximum surface brightness versus magnitude}

The larger the distance, the harder it becomes to distinguish the image of a GC from that of a point source, which may be a star or a small background object. We will thus fall back to deriving a catalogue of candidates for subsequent follow-up. Catalogue column combinations that inform indirectly about source compactness make it possible to reduce the number of candidates drastically \citep{Cantiello2018}. For example in \citet{Voggel2020} the astrometric excess noise tabulated in Gaia catalogues was combined with magnitude around Centaurus A, and a source catalogue of several million point-like sources was reduced to several hundred good candidates, which then facilitated a much more focused follow-up. In the central 30\,kpc of Cen\,A,  the spectroscopic success rate of that selection was 68\% and outside of 30\,kpc it dropped to 8\% as in the sparse very outer regions real GCs are more rare \citep{Hughes2021}.

We test the purity and completeness of a GC catalogue based on simple selection criteria from the \Euclid photometric catalogue that is provided with each observation. As this is an idealised framework based on simulations with limited GC sizes and magnitudes, and only a single type of host galaxy, we assume that in reality the foreground/background separation from the GCs will be somewhat more difficult than what we test here. Nevertheless, this offers us a first estimate of the expected contamination by fore- and background objects.

\begin{figure}
\centering
\includegraphics[width=\hsize]{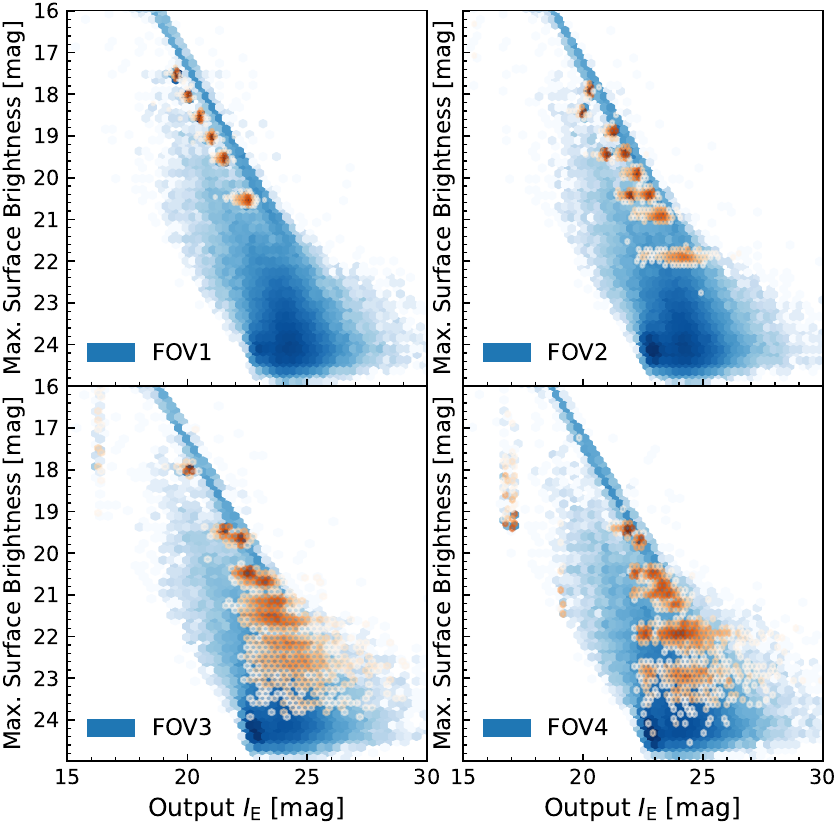}
\caption{Maximum surface brightness of all sources in the \Euclid output catalogue of each of our four simulated tiles as a function of the $\IE$ magnitude. All sources are plotted as the blue density distribution and the GCs that we added to the simulations are shown as the orange density distribution.} 
\label{fig:mumax}
\end{figure}

We plot the maximum surface brightness measured by the \Euclid\ pipeline on the VIS image of a GC against the ouput \IE\ magnitude in Fig.~\ref{fig:mumax}, for the four simulated {\tt FoV}. We plot all the sources of a given {\tt FoV} in blue and the injected artificial GCs in orange. The discrete distribution of GCs in this figure derives from the fact that there are only six input GC magnitudes and two GC sizes in the simulation. The objects that fall into a straight vertical line at $\IE=16$--17 are the nuclear star clusters that we placed at the centre of each galaxy. Their magnitude is heavily affected by the host galaxy and they are not well visible in FoV1 and FoV2 because there are only four and 16 galaxies and thus nuclei in each of these FoVs. 
The surface brightness comparison can be useful to identify GCs, as the maximum surface brightness of GCs is in theory lower than those of stars, but higher than for galaxies. The figure lets us evaluate how much that effect is present in simulated \Euclid data.

Figure~\ref{fig:mumax} shows that the GCs fall into a specific parameter space that is offset from the locus of galaxies and stars. Stars form a tight sequence above the GCs location as they have a higher surface brightness compared to GCs, and galaxies form a cloud of data points at lower surface brightness. This separation of objects is clearest for the closest {\tt FoV1} at 5\,Mpc and with increasing distance in {\tt FoV4} the GCs do not separate out easily from background galaxies anymore. Their location in terms of surface brightness has strong overlap with galaxies and stars, and thus more parameters are needed to ensure an efficient pre-selection of GC candidates.

\section{Globular clusters in early \Euclid data}

\begin{figure}
\centering
\includegraphics[width=\hsize]{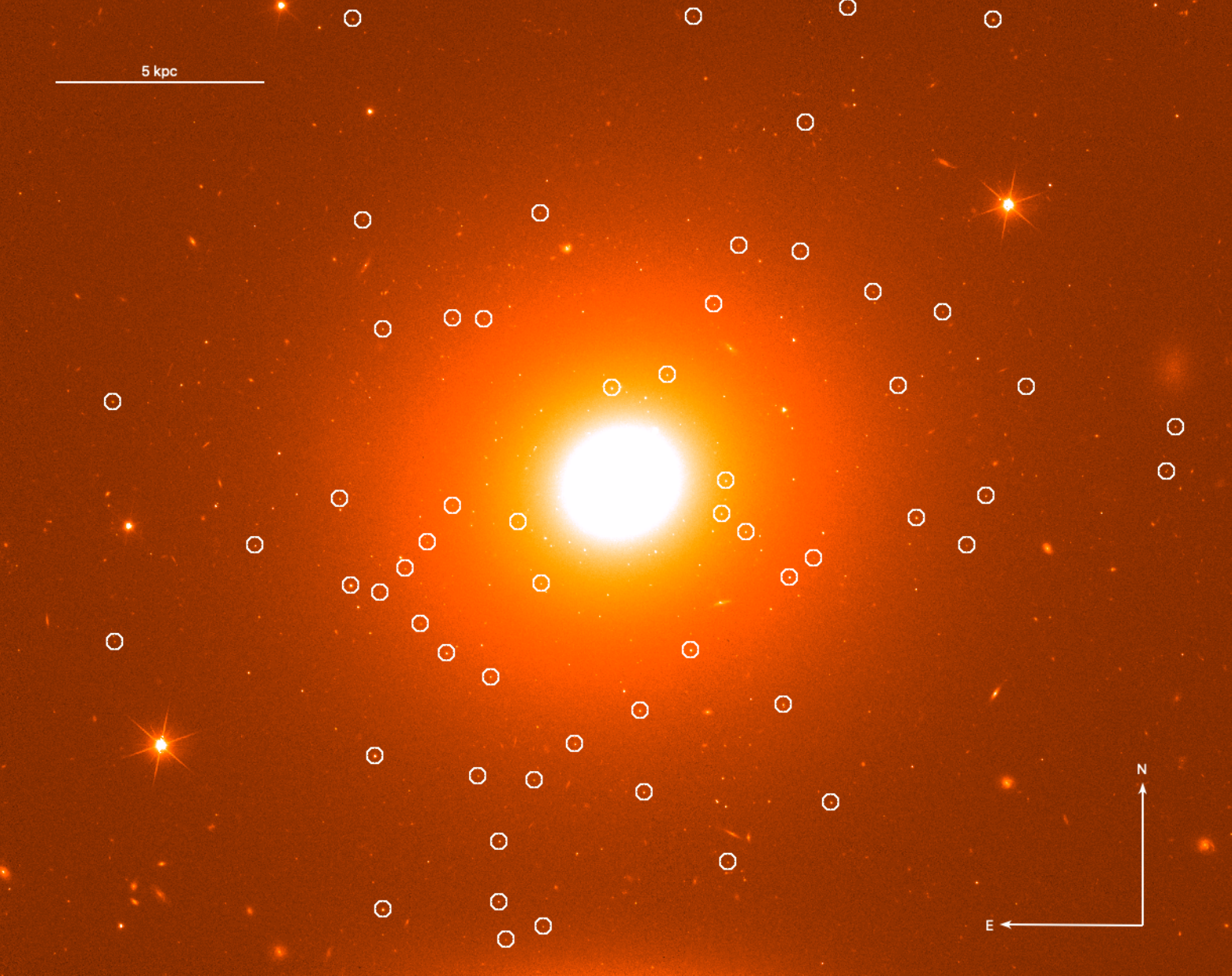}
\caption{Cut-out of the \Euclid VIS image focused on NGC\,1374. The circles mark the location of the GC candidates in this region. } 
\label{fig:NGC1374}
\end{figure}

We can test the combination of selection methods further using \Euclid early-release observations (ERO) of the Fornax galaxy cluster \citep{EROFornaxGCs, EROData}, which have been acquired recently. The Fornax cluster was chosen because, being located at a distance of about 20\,Mpc, it is one of the nearest galaxy clusters to the Milky Way, and it contains a large number of GC-hosting galaxies. 
Two VIS observations with an exposure time of 560\,s each were taken, centred on position RA=\ang{03;36;08.759} Dec=\ang{-35;16;00.38}, and were co-added and processed with an ERO-specific pipeline, which is half of the expected observing time of the EWS where four exposures are the standard. A cut out of the image focussed on NGC\,1374 is shown in Fig.~\ref{fig:NGC1374}. The co-added full image is shown in the bottom panel of Fig.~\ref{fig:gcdensity}. 

\subsection{Applying a Gaussian-mixture model to select GC candidates in Fornax}
\begin{figure}[ht!]
\centering
\includegraphics[width=\hsize]{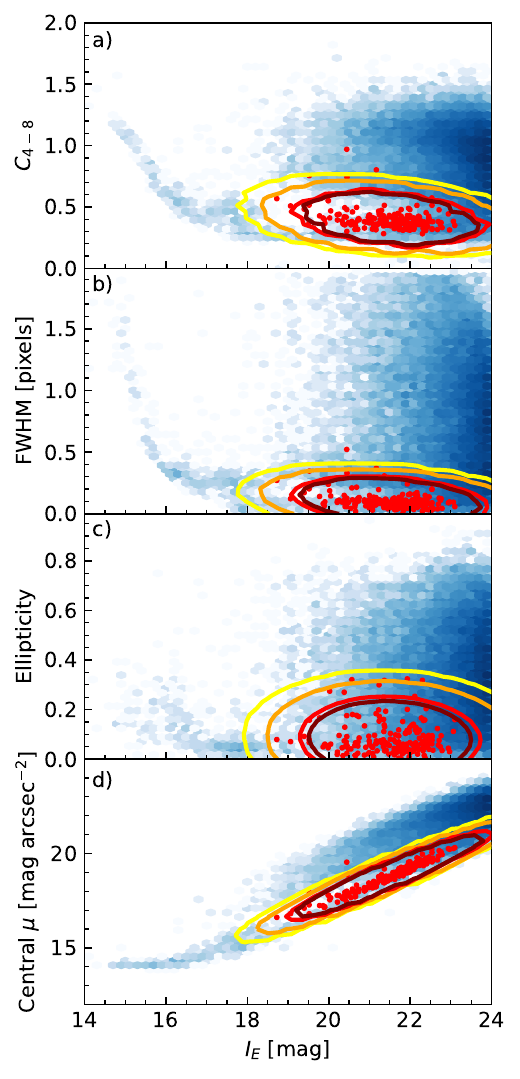}
\caption{Parameters of detected sources against VIS $\IE$ magnitude. Panel a) is the concentration index $C_{4-8}$, which is the difference in magnitude between the aperture with a radius of 4 and 8 pixels. The parameter shown in panel b) is the FWHM of the sources in arcsec, panel c) is the ellipticity and panel d) is the maximum surface brightness. All four parameters are derived from the ERO-Fornax catalogue and are intrinsic size parameters deconvolved with the PSF. Known GCs are shown as red symbols. The blue shaded density distribution represents all photometric sources in the Fornax FoV. The yellow, orange, red and maroon ellipses mark the levels that include 95\%, 90\%, 68\%, and 50\% of all known GCs in the Gaussian mixture output. } 
\label{fig:Gauss_fornax}
\end{figure}  

Follow-up observations of previous ground- and space-based surveys of Fornax have established a set of spectroscopically confirmed GCs  \citep{Saif2021b,Chaturvedi2022}, that we can use to train an automatic classification algorithm. We initially intended to use \Euclid-pipeline catalogues for the Fornax VIS data to ensure comparability with the results from the simulations, but this was not possible so early after launch with data taken in a sequence that does not correspond to the standard sequence of the future EWS. We thus ran {\tt Source Extractor} \citep{Bertin1996} on the VIS images independently of the pipeline. The resulting catalogue is called the `ERO-Fornax catalogue' hereafter. 

When matching the known GCs with the ERO-Fornax catalogue, we found a handful of sources with ellipticity $e>0.6$ and $\rm{FWHM}>0.75$\,pixels in the VIS images. Typical GCs are round and do not have large half-light radii. We clipped our sample at 3$\sigma$ above the mean in FWHM and ellipticity, which resulted in a sample of 181 GCs with ellipticities smaller than $0.38$ and $\rm{FWHM}< 0.64$\,pixels. Upon visual inspection, we found that the rejected sources were affected by blending with faint nearby sources, although this did not trigger the blending flag in {\tt Source Extractor} as these blending sources were too close and faint to be picked up as secondary sources. In two of these objects they were close to the centre of larger galaxies, which caused the faulty ellipticities. Sources that are blended will be challenging cases to identify as bona fide GCs with such automated classification methods.

The ERO-Fornax catalogue provides us with magnitudes in apertures of different sizes and basic size measurements such as the FWHM for all detected sources in the field. More details on this catalogue are given in \cite{EROFornaxGCs}. The goal was to identify a parameter subspace in which a good classification could be achieved, thus distinguishing between the different classes of objects. We use the $\IE$ magnitude, FWHM, ellipticity, and central surface-brightness parameters, which are comparable to the parameters used in the simulation test. The ERO-Fornax catalogue contains no concentration parameter. We thus constructed a new one that is the difference between magnitudes in an 4-pixel and a 8-pixel aperture, labelled $C_{4-8}$.
In Fig.~\ref{fig:Gauss_fornax} we show the distribution of our training set of known GCs (red) in comparison to the full ERO-Fornax catalogue (blue density bins). 
Although the \Euclid data set extends to $26$ magnitude, we decided to limit our data set for classification to $\IE \leq 24$ because that is the magnitude limit of our training sample. 

\begin{figure}
\centering
\includegraphics[width=\hsize]{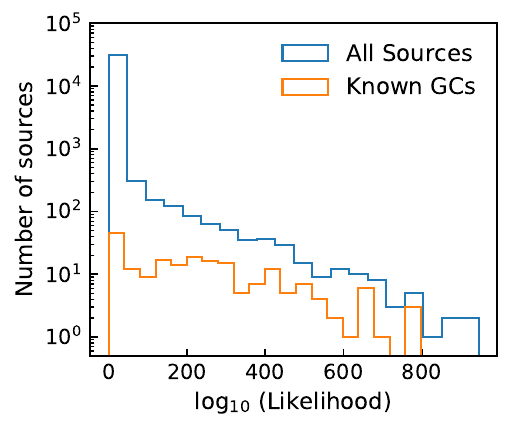}
\caption{Non-normalised distribution of log-likelihoods of extracted sources: known GCs (orange) and all sources (blue). The vast majority of the source catalogue falls in the smallest bin of probability, which includes all log-likelihoods below 40 in the units of this figure. }
\label{fig:likelihood}
\end{figure}

We train a Gaussian mixture model on the known GCs in the five-dimensional parameter space defined above. The results of this training is shown as ellipses in Fig.~\ref{fig:Gauss_fornax}. The red and yellow ellipses mark the levels that include 50\%, 90\%, and 95\% of all known GCs in the Gaussian mixture output. This Gaussian model, which is essentially a description of the region of five-dimensional parameter space accessible to GCs, is then applied to the whole photometric data set of 32\,554 sources brighter than 24\,mag whose nature is unknown. Sources near the centre of the Gaussian model are intrinsically more likely to be GCs than sources near the outer contours. The corresponding likelihood distribution of all sources is shown in Fig.~\ref{fig:likelihood}. At the cost of lower completeness, one could consider additional truncations based on the preferential locations of non-GC populations, but we focus on completeness here.

The vast majority of the catalogue sources have very small likelihood values and are thus easily excluded as GCs. The tail at high likelihoods for the full catalogue is where the most likely GCs are located, with properties similar to those of known objects. We make a cut at a log-likelihood of 8.5 (in the units of Fig.\,\ref{fig:likelihood}) for objects that we consider good candidates. That value corresponds to the ellipse that includes 90\% of all known GCs (second ellipse from the outside in Fig.~\ref{fig:Gauss_fornax}) assuring a good completeness of our candidate sample. This results in 1541 sources, out of 32\,363 in total, that we consider as good GC candidates. The known objects have been removed from the source catalogue so these are all new candidates. While we do expect a substantial contamination of faint galaxies and stars, the decrease of our original source catalogue by 95\% is the key for an efficient selection of GC candidates solely from the photometric catalogues of the full EWS. 
 
\begin{figure}
\centering
\includegraphics[width=\hsize]{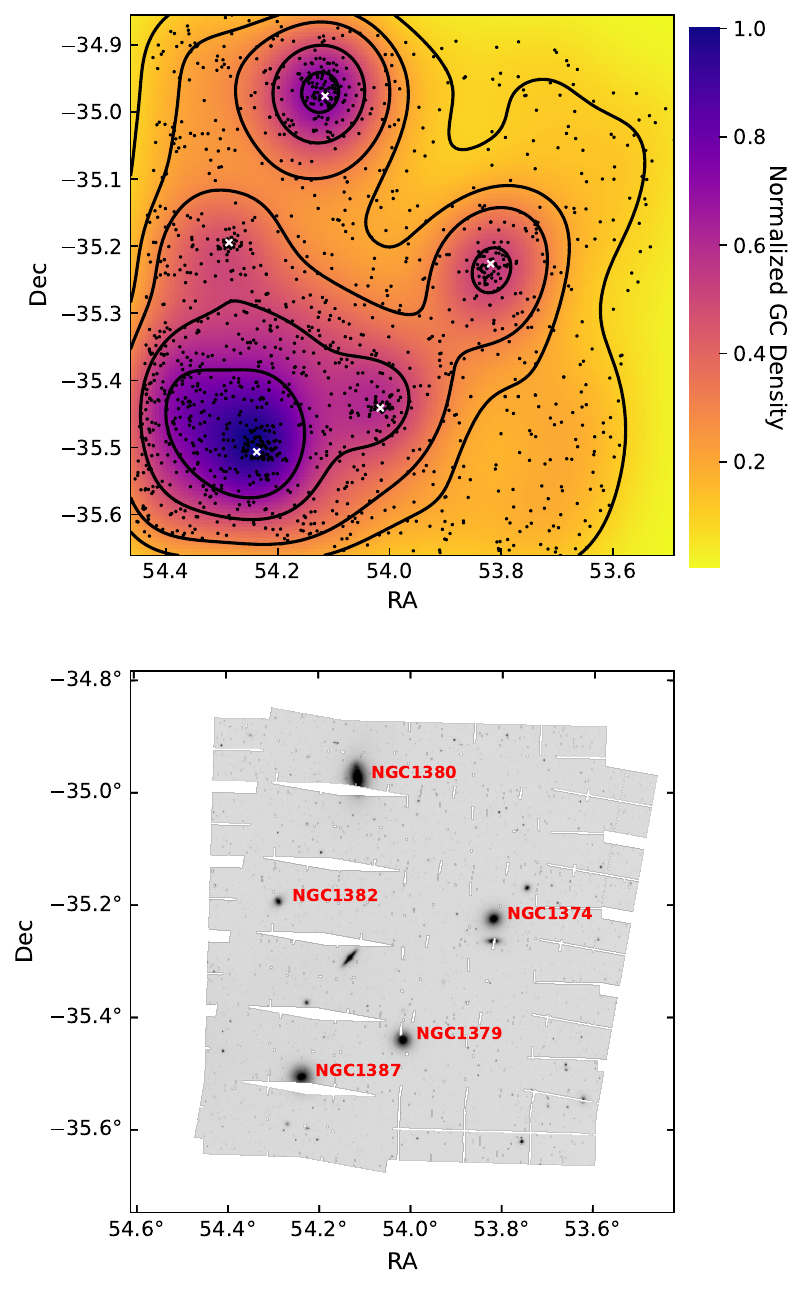}
\caption{Spatial distribution of GC candidates. \textit{Top panel:} Positions of the 1541 high likelihood GC candidates are shown as black data points and location of the main galaxies is shown as white crosses. The colour and contour shows the normalised kernel density estimate of GC density. \textit{Bottom panel:} The \Euclid image from the VIS instrument in the $\IE$ band. The white locations are where there are gaps in the data. } 
\label{fig:gcdensity}
\end{figure}

The spatial distribution of the 1541 high-likelihood GC candidates is shown in the top panel of Fig.~\ref{fig:gcdensity}. The colour and contours in the top panel are the normalised kernel density estimate for the GCs using a Gaussian kernel. 

The density of GC candidates peaks at the location of the main galaxies in the Fornax FoV (marked with a white cross). This is consistent with what we expect from typical GC systems, which display a radially decreasing number density profile centred on their host galaxy. This suggests that our automatic selection method is picking up a large number of real GCs. From this figure it appears that there is a bridge of overdensity of GCs between the galaxies going from NGC\,1387 at the bottom left of the FoV to the galaxy on the very top (NGC\,1380) as well as to the galaxies on the right side of the FoV (NGC\,1374 and NGC\,1379). However this impression of a bridge of intracluster GCs could also be caused by a projection effect of the GC systems of the two smaller galaxies along the path of these two `bridges'. Three such intracluster GC bridges were found in the Fornax deep survey \citep{Cantiello2020}, however our FoV does not cover those three previously known GC bridges. A more detailed analysis of the intracluster GCs and a comparison to the intracluster light is presented in \cite{EROFornaxGCs}. With the full \Euclid survey data, such bridge-candidates will be detectable in all nearby galaxy clusters.

The above analysis shows the efficiency of the training method in a field with known GCs. For nearby galaxies and clusters this will enable us to derive a candidate catalogue with a high completeness based solely on catalogue data. Such a predictive modelling has the advantage that it is computationally cheap and fast, thus enabling an efficient selection of GC candidates for further study. We do expect significant contamination from both foreground and background sources, but the removal of typically more than 90\% of contaminants will enable us to make a very efficient follow-up while keeping a large completeness of GCs in our sample. 

This method is also adaptable to fields where we have no good training sets. Using the expected FWHM and magnitude of GCs at a given distance one can use a method similar to a `matched filter' where the location of GCs can be modeled given an array of possible distances. These expected parameter spaces can then be used to provide the best GC candidates even in a field where we do not know the distances of potential GCs. Here we investigated, how detecting GCs in Fornax could be scaled to the real \Euclid data. This simple and computationally efficient method is able to reduce background contamination by at least 90\%, indicating that such a machine learning analysis based on GMM is a very promising approach

\section{Size measurement for globular clusters in VIS images}\label{sizes}

The $\ang{;;0.14}$ FWHM of the VIS point-source images allows us to carry out size measurements for GCs in nearby galaxies. GC sizes are useful to establish their membership of the host galaxy and also to measure distances \citep{Jordan2005, Masters2010}. We take the catalogue of 233 spectroscopically confirmed GCs from \cite{Saif2021b} that fall within the coverage area of our Fornax VIS observations and use these as a validation sample for size measurements based on light-profile fitting. 
These GCs are all brighter than $r=23$ with their magnitude distribution peaking at $r=21.5$ and the brightest GC is at $r=18.7$.
As a reference we also fit known stars, which are true point sources. The two codes used are \texttt{Ishape} \citep{Larsen1999} and \texttt{GALFIT} \citep{Peng2002}.
Both fit two-dimensional analytic profiles convolved with the PSF to the image of a source, provide the best-fit parameters and allow the user to examine model-subtraction residuals. 

First, we create postage stamp cut-outs of $300\,\textrm{pixels}\times300\,\textrm{pixels}$ centred on each known GC  that will then be fed into \texttt{GALFIT} \citep{Peng2002}. \texttt{GALFIT} also requires an input PSF with which the models are convolved. We derive this PSF by oversampling and stacking the bright and non-saturated point sources using {\tt SWarp} \citep{swarp}.
The point sources for the modelling were selected to be non-saturated and to fall well within the magnitude, FWHM, ellipticity, and classifier parameter range of typical stars. More details on this PSF creation are given in \cite{EROFornaxGCs}. We use the same PSF for both papers.

We then fit each of the known GCs with a King light profile \citep{King1962}. We use a standard King profile and fit for the core radius $r_{\rm c}$ and the concentration $c := r_{\rm t}/r_{\rm c}$, where the tidal radius $r_{\rm t}$ is set to $r_{\rm t}=\ang{;;1.2}$. Fixing the latter is necessary because the faint truncation radius is not well constrained by the data, a free search often causing \texttt{GALFIT} to fail to converge. The results for $r_{\rm c}$ are not very sensitive to the choice of $r_{\rm t}$ among values compatible with normal GC or UCD properties. Adopting a distance of $20.9$\,Mpc to Fornax \citep{Blakeslee2009}, $r_{\rm t}=\ang{;;1.2}$ corresponds to a core radius of $4$ or $5$\,pc when $c=25$ or $c=30$, or an FWHM of $8$ to $10$\,pc ($\ang{;;0.08}$ to $\ang{;;0.1}$), which is representative of a large GC or a small UCD.

All other parameters in the fitting procedure such as position, central surface brightness, core radius, and the axis ratio are allowed to vary. We perform a similar fitting of light profiles of the confirmed GCs using \texttt{Ishape} as a comparison method. \texttt{Ishape} also needs a PSF, which is convolved with the analytic King model \citep{King1962} until the best fit to the data is obtained.
The PSF was derived by a similar procedure as described above, and then oversampled by a factor of 10 as required by \texttt{Ishape.}
In the \texttt{Ishape} run only circular light profiles were allowed.
The \texttt{Ishape} fits also assume a King profile, in this case with a fixed concentration parameter of $c =30$.

We also select known stars from Gaia DR3 in our field of view to make a comparable measurement for them and use them as a baseline for a non-extended source. For this, we select every Gaia source in the ERO field-of view that had parallax and proper motion in both RA and Dec direction detected at higher than 3\,$\sigma$. We also require that they are fainter than 19th magnitude (in Gaia $G$) to avoid possible saturation in the \Euclid VIS band, which occurs at around 18th magnitude.

\begin{figure}
\centering
\includegraphics[width=\hsize]{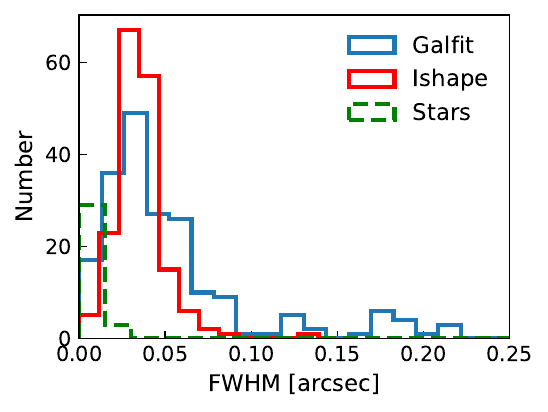}
\caption{Size distribution of known GCs in the \Euclid imaging. The blue histogram shows the results from \texttt{GALFIT} and in red those from \texttt{Ishape}. The comparison data set of known stars is shown as dashed green histogram.} 
\label{fig:fwhm}
\end{figure}

Figure~\ref{fig:fwhm} shows the distribution of estimated intrinsic FWHMs of the three source categories: known GCs measured with \texttt{GALFIT} (blue) and with \texttt{Ishape} (red), and known stars (green dashed).
The stars selected with Gaia are unresolved as expected, with estimated {\em intrinsic} FWHMs smaller than $\ang{;;0.03}$. The spectroscopically confirmed GCs or UCDs are mostly extended, with estimated intrinsic FWHMs between $\ang{;;0.03}$ and $\ang{;;0.1}$. This corresponds to a physical size of $3~{\text{pc}} < \text{FWHM} < 10~{\text{pc}}$, which is equivalent to half-light radii of  $4~{\text{pc}} < r_{\text{h}} < 15~{\text{pc}}$ (with some dependence on the value of $c$). If for simplicity we consider that normal GCs have half-light radii between 1 and 10\,pc while very bright GCs or UCDs have larger radii between 10 and 50\,pc \citep{Jordan2005,Liu2020}, this shows that we do resolve the half-light radii of all the compact objects but the smallest GCs.

\begin{figure}
\centering
\includegraphics[width=\hsize]{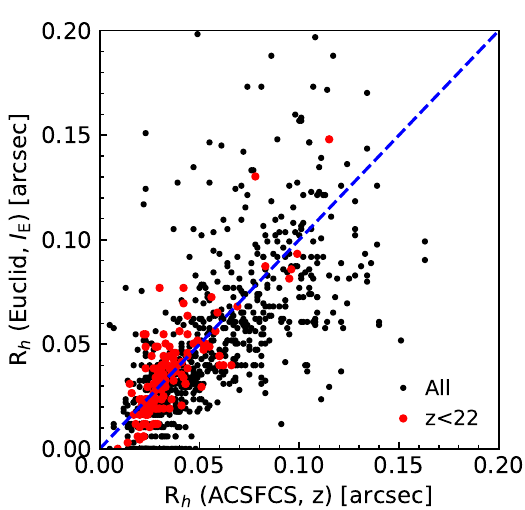}
\caption{Globular cluster half-light radii measured with \texttt{Ishape} on \Euclid VIS images versus measurements from the ACS Fornax Cluster Survey \citep{Jordan2015}.} 
\label{fig:Rh_ACS_VIS}
\end{figure}

To further validate our size measurements on the VIS images, we compare with the catalogue of GC properties from the ACS Fornax Cluster Survey, ACSFCS \citep{Jordan2015}. In that work, GC half-light radii were measured on HST/ACS images by fitting \citet{King1966} models, convolved with the PSF, to the ACS images.
As HST has four times the effective area of \Euclid and the ACS images are sampled at $\ang{;;0.05}$, these data provide a useful reference against which our measurements can be verified. The ACSFCS catalogue contains sources to a limiting magnitude of $z\simeq25$\,[AB].
The result of this comparison is shown in Fig.~\ref{fig:Rh_ACS_VIS} for the ACSFCS $z$-band measurements. The FWHM values measured by \texttt{Ishape} have been converted to half-light radii adopting the $\ang{;;0.1}$ pixel scale of the VIS instrument and the conversion $r_{\rm h} = 1.48 \times \mathrm{FWHM}$ as appropriate for a King $c=30$ profile \citep{Larsen1999}. The full set of sources contained in the ACSFCS catalogue that also have \Euclid size measurements are shown with black dots, while sources brighter than $z=22$ are shown with red dots. Overall, the two independent size measurements show a strong correlation and follow the 1:1 relation (indicated by the dashed line) quite well. The standard deviation of the difference between the two sets of measurements is $\sigma_\mathrm{ACS-Euclid} = \ang{;;0.024}$
(2.2\,pc) for the full sample and $\sigma_\mathrm{ACS-Euclid} = \ang{;;0.013}$  (1.2\,pc) for $z<22$. Again, this shows that with careful modelling of the PSF, it is possible to measure sizes for objects down to a small fraction of the size of a pixel. 

This first size analysis shows that with \Euclid imaging we can obtain a measurement of GC sizes at the distance of Fornax ($\simeq$\,20\,Mpc). It is clear that accurate modelling of the PSF will be crucial to the classification and measurement of such compact sources. Sampling effects may also play a role in order to preserve information at the sub-pixel level. Because our Fornax VIS mosaic is based on just two images instead of the four that would be available in the standard Euclid Wide Survey, we postpone the analysis of other recovered model parameters to future work.

\section{Conclusion}\label{sec:discussion}
In this work, we have tested how well GCs can be detected in \Euclid\ VIS imaging in both simulated pre-launch images and the first early-release observations of the Fornax cluster. We aimed to study the potential and capabilities of the Euclid Wide Survey for the discovery and study of GCs in general. Our main findings are as follows.

Based on empirical scaling relations that relate GC numbers to magnitude, together with a catalogue of known galaxies within 100\,Mpc that fall in the  $14\,000$\,\rm deg$^{2}$ footprint of the Euclid Wide Survey, we expect these local-Universe galaxies to host around 830\,000 GCs of which about 350\,000 are within the surveys detection limits of the \IE\ band. For about half of these GCs we predict that three infrared colours in the \YE, \JE, and \HE\ bands will be available as well.

These numbers provide a first prediction, but the final number of GCs could be lower depending on our ability to correct for contamination from the light of the host galaxy or for blending with foreground and background sources. However, these losses might be compensated by the fact that not every galaxy within the local Universe is included in the HECATE catalogue that was our main reference, due to incompleteness at the faint end of the galaxy luminosity function. The GCs in these missing dwarf galaxies will add to those estimated GC numbers. Furthermore, this estimate does not include any intracluster or intragroup GCs, whose numbers in relation to galaxy-cluster mass is not well understood. Our study shows that \Euclid\ will be able to provide an estimate of the intracluster GC abundance in local-Universe galaxy clusters. Thus overall it is likely that the quoted numbers of detectable GCs are only a lower limit and the true numbers are even higher.

The typical magnitude limit of the Euclid Wide Survey in the VIS images is at $\IE=26.2$, which means that at 50\,Mpc around 50\% of a typical GC system is within the detection limit. Therefore for any galaxy within that distance the brighter half of its GC luminosity function will detectable by \Euclid. At distances below 50\,Mpc we cover the peak of the GC luminosity function whose peak can be used as a distance measure (e.g., \citealt{Whitmore1997}). In \citep{EROFornaxGCs} the GC luminosity function and its peak magnitude was well determined in Fornax cluster galaxies based on GC candidates only, thus serving as a rough distance estimator.

GCs will also be detected in the infrared \YE, \JE, and \HE\ bands down to a magnitude limit of $\IE=24.5$. Infrared colors are crucial to study a GCs stellar populations and to help distinguish them from fore- and background objects. This magnitude limit implies that the brightest half of a GC system of a given galaxy is detectable in the infrared out to 30\,Mpc of distance.

The detectability of GCs depends strongly on their contrast against the residual surface brightness surrounding it. For galaxies with smooth profiles, we find that a nearly full recovery fraction of 80 to 90\,\% is expected at a residual host galaxy surface brightness of $25\,\text{mag}\,\text{arcsec}^{-2}$ in \IE, which is independent of distance. This surface brightness limit corresponds to 5 times the effective radius of the host galaxy at a distance of 20\,Mpc when no galaxy light is removed. With a careful removal of the host galaxy light, GCs can be detected much closer to the centre of the galaxy and thus the removal of galaxy light is crucial for the detectability of GCs. As removal of host galaxy light is costly, we predict that Euclid will have much more complete detection of GCs beyond the effective radii of host galaxies.

We used early \Euclid\ data on the Fornax cluster to test our ability to select candidate GCs from a photometric catalogue. We have successfully trained a mixture model on the location of spectroscopically confirmed GCs in the photometric catalogue of the observations. We assigned a likelihood to be a bona fide GC depending on their location in a five-dimensional parameter space that includes magnitude, surface brightness, and size as well as a parameter for concentration. Making a reasonable likelihood cut that ensures 90\% completeness in our training set, we derive a set of about 1500 new GC candidates within the Fornax cluster field observed by \Euclid. This selection has reduced the initial photometric source catalogue by 95\%. The method is computationally efficient as it only requires a photometric catalogue of the \Euclid observations and an expected parameter space of GCs at a given distance as input parameters. This is essential as a full re-analysis of all \Euclid images to study GCs is computationally prohibitive. Therefore, we plan to use a strategy akin of a matched-filter method where we comb the \Euclid photometry for a given set of expected GC properties at each of several distances for a first selection of GC candidates.

We test how well GCs are spatially resolved in early \Euclid images of the Fornax cluster. Typically, in space-based images, light profile fitting was used to determine which objects are bona fide GCs, and sometimes also to establish the cluster membership of their host galaxies as well as to measure their distances. With \Euclid this is possible as well, however the exact limits for which GCs will appear as extended are hard to estimate from empirical scaling relations and simulations alone. With the new first release \Euclid data of Fornax we have measured the sizes of known GCs in the cluster with two different methods. For objects that we can clearly distinguish from stars, we find sizes of  $4\,{\rm pc} < r_{\rm h} < 15\,{\rm pc}$, that correlate well with {\em Hubble} Space Telescope sizes where these are available.

This demonstrates that the sizes we measure for those GCs are true extensions. The typical half-light radius of GCs in the Universe is close to 2\,pc, and is equivalent to only 0.2\,pixels in the VIS imager at the distance of Fornax. This is in the grey zone in which the present exploration, which uses a non-optimal mosaic of only two VIS images and a single PSF-model, does not always provide a clear separation from stars. The better signal-to-noise ratio of standard survey images and a more accurate modelling of the PSF will be crucial to the classification and measurement of such very compact sources.  
However, this result shows that we can indeed measure the sizes and light profiles of most GCs at $d<20$\,Mpc, and once the full \Euclid survey is available, the improvements in the modelling of the PSF with increasing experience of the data will improve this limit further.

\section{Outlook}\label{sec:conclusion}

In this paper we have shown the potential that \Euclid data will present for advancing globular cluster science in the future. We estimate that there are $350\,000\pm5000$ GCs that are possibly detectable in the \Euclid \IE band. This value is based on taking into account all known galaxies within 100\,Mpc in the Euclid Wide Survey area, and on estimating their total number of GCs and their magnitude distribution. This number is an estimate on how many GCs should be visible in \Euclid VIS images. The real number will likely be higher due to the number of intracluster GCs that were not included in the simulation as well as the faint-end incompleteness of our input galaxy catalogue. This increase might be offset by a reduction in detected GCs due to host-galaxy background contamination, blending, and other observational effects. 

We also expect high-resolution VIS-imaging data for hundreds of thousands of GCs and detailed infrared colours for about half of those enabling us to produce the largest homogenous GC candidate catalogue in an non-spatially biased survey area over a third of the sky. Such a catalogue will contain many newly detected GCs, especially in the distant outskirts of galaxies and in the intracluster areas and allow for a census of GC colours, positions, luminosity function and other properties as a function of host galaxy. Such a catalogue will also allow for targeted spectroscopic follow-up of GCs that makes them usable as dynamical tracers of galaxy mass profiles and the dynamical state of entire galaxy clusters.

Using the first real \Euclid data on the Fornax cluster we found that the generation of GC candidate catalogues using machine learning is an efficient method to reduce the initial contaminants from the source catalogue by 95\%, while keeping the completeness above 90\%. This efficient reduction into a candidate catalogue allows more efficient targeted follow-up studies of the light profiles as well as spectroscopic follow-up and confirmation of bonafide GCs. Such automatic classification is extremely computationally efficient and can be easily expanded to a blind search across the whole Euclid Wide Survey footprint by using the expected parameter range of GCs depending on their distances. In the future we plan to employ a matched filter technique where we find GCs for their expected parameter as a function of distance. Such a matched filter run with a whole range of distances can provide GC candidates even for galaxies where we have no estimate of their distance. Such an efficient selection method is essential as the data volume of \Euclid images is much too large to use standard manual GC search methods.

Having used the Fornax cluster as a benchmark, we confirm that individual GCs are spatially more extended than pure point sources at 20\,Mpc. This knowledge shows that a catalogue of GC light profiles and structural parameters will be possible for all sufficiently bright GCs within within 20\,Mpc in the future \Euclid data. 
Our analysis of both simulated and first real data shows the vast number of GCs that will be present in \Euclid image and will be spatially resolved and have infrared colours. We expect \Euclid to be revolutionary to GC science by increasing the number of GCs accessible with space-quality high-resolution imaging by an order of magnitude or more. In particular, the wide field coverage of 14\,500\,deg$^{2}$ enables us to search for GCs in a way that is not biased towards the centres of galaxies. We especially expect a lot of new discoveries in the outskirts of galaxies and for intracluster GCs, which have been not studied with space telescopes that have a small field-of-view.


%
%

\begin{acknowledgements}
This work was supported in part by Agence Nationale de la Recherche (France) under grant ANR-19-CE31-0022, 
by project `INAF Exploration of Diffuse Galaxies with Euclid' (INAF-EDGE, Italy, 2022, P.I. Leslie K. Hunt)
and by grant `LEMON' (INAF, Italy, 2022).
\AckEC
\AckERO 
Co-funded by the European Union. Views and opinions expressed are however those of the author(s) only and do not necessarily reflect those of the European Union. Neither the European Union nor the granting authority can be held responsible for them. JHK acknowledges grant PID2022-136505NB-I00 funded by MCIN/AEI/10.13039/501100011033 and EU, ERDF.
\end{acknowledgements}
%
%

\bibliography{Euclid}
\label{LastPage}

%
%

  


\end{document}